\documentclass[12pt]{article}
\usepackage{amssymb,amsmath,epsfig,mathtools,caption}

\begin{document}

\title{\bf Tidal Forces in Kiselev Black Hole}
\author{M.Umair Shahzad$^{1}$ \thanks{m.u.shahzad@ucp.edu.pk} and
Abdul Jawad$^2$ \thanks{jawadab181@yahoo.com; abduljawad@ciitlahore.edu.pk}\\
\small $^1$CAMS, UCP Business School, University of Central Punjab,\\
\small  Lahore, Pakistan.\\
\small $^2$Department of Mathematics, COMSATS Institute of Information\\
\small Technology, Lahore-54000, Pakistan.\\}

\date{}
\maketitle

\begin{abstract}

The aim of this paper is to examine the tidal forces occurred in
Kiselev black hole surrounded by radiation and dust fluids. It is
noted that radial and angular component of tidal force change the
sign between event and Cauchy horizons. We solve the geodesic
deviation equation for radially free falling bodies toward Kiselev
black holes. We explain the geodesic deviation vector graphically
and point out the location of event and Cauchy horizons in it for
specific values of radiation and dust parameter.
\end{abstract}

\section{Introduction}

At present, type 1a supernova \cite{1}, Cosmic microwave background
(CMB) radiation \cite{2} and large scale structure \cite{3,4} have
shown that our universe is currently in accelerating expansion
period. Dark energy is responsible for this acceleration and it has
strange property that violates the null energy condition (NEC) and
weak energy condition (WEC) \cite{5,6} and produces strong repulsive
gravitational effects. Recent observations suggests that
approximately 74\% of our universe is occupied by dark energy and
the rest 22\% and 4\% is of dark matter and ordinary matter
respectively. Nowadays dark energy is the most challenging problem
in astrophysics. Many theories have been proposed to handle this
important problem in last two decade.

With the discovery of cosmic acceleration, black holes (BHs)
phenomenon have become the most fascinating in illustrating their
significant physical properties. There exists two major types of
vacuum BH solutions in general relativity, i.e., uncharged (for
example Schwarzschild BH) and charged (for example
Reissner-Nordstrom BH). These BHs have been thoroughly investigated
by many authors over the years. For example, there exists a
well-known phenomena in which a body experiences compression in
angular direction and stretching in radial direction when it falls
toward the event horizon of uncharged static BHs
\cite{16,17,18,19,20}. However, in Reissner-Nordstrom BH, a body may
experience stretching in radial direction and compression in angular
direction depends upon two phenomenons $1)$ the location of body and
$2)$ charge to mass ratio of BH \cite{21}. Tidal forces change their
sign in radial or angular direction at certain points of
Reissner-Nordstrom BH unlike Schwarzschild BH. Geodesics deviation
of Schwarzschild and Reissner-Nordstrom space-times are studied in
detail by \cite{15,16,21,22,23}. Ghosh and Kerr \cite{24} discussed
geodesics deviation and geodesic motion in wrapped space-time with
extra dimension of time. It is also realized that the geodesics
deviation needs to analyze using full general relativity on the
other hand tidal forces can be found with Newtonian mechanics if an
extra force coming from general relativity is added.

However, several BHs of Einstein general relativity in non-vacuum
case have also been presented \cite{7,8,9,10,101,11} which need more
physical illustrations. One of them is Kiselev BH which possesses
new set of phenomena unlike Schwarzschild BH because of the
important complex properties \cite{11} and this BH has been
surrounded by various types of matter depending on state parameter
$\omega$. Kiselev BH have been investigated through various
phenomenon, i.e., accretion \cite{12}, strong gravitational lensing
\cite{13}, thermodynamics and phase transition \cite{14}. In this
work, we apply the technique of \cite{29} on the solution of Kiselev
BH surrounded by energy matter i.e. we consider Kiselev BH
surrounded by dust and radiation parameter derived by Kiselev
\cite{11}, in which we consider non zero electric charge, dust and
radiation parameter but no angular momentum. They are exact
solutions of Einstein Maxwell equation \cite{15}, in the case of
vanishing dust and radiation parameter it reduces to RN space time
and in the case of vanishing electric charge it reduces to SH space
time.

In this paper we discuss the tidal forces in Kiselev space-time and
consider its two cases which leads to Kiselev space-time surrounded
by dust $(\omega = 0)$ and radiation $(\omega = 1/3)$. We solve the
geodesic deviation equations to observe the variation of test body
in-falling radially toward the Kiselev BH for specific choices of
dust and radiation parameter. This paper is organized as follows: In
Sect. \textbf{2}, we discuss Kiselev BHs, its two special cases and
radial geodesics. In Sect. \textbf{3}, we derive the tidal forces in
Kiselev space-time on a neutral body in radial free fall. In Sect.
\textbf{4}, we find the solutions of the geodesic equations in
Kiselev space-time. In the end, we conclude our results. In this
paper, we use the metric signature (+, -, -, -) and set the speed of
light $c$ and Newtonian gravitational constant $G$ to $1$.

\section{Kiselev black holes and its two special cases}
The line element of static charged BH surrounded by energy-matter is
given by
\begin{equation}
ds^{2}=f(r)dt^{2}-f(r)^{-1}dr^{2}-
r^2({d\theta}^{2}+\sin\theta{d\phi}^{2}),\label{1}\end{equation}
with
\begin{equation}\label{2}
f(r)= 1-\frac{2M}{r}+\frac{q^2}{r^2}-\frac{\sigma}{r^{3\omega+1}},
\end{equation}
where $M$ and $q$ are the mass and electric charge, $\sigma$ and
$\omega$ are normalization parameter and state parameter of matter
around BH, respectively \cite{11}. We assume $\omega=1/3$ which
becomes Kiselev BH surrounded by radiation and $\omega=0$ Kiselev BH
surrounded by dust. For Kiselev BH surrounded by radiation, the
radial coordinated of horizons are obtained by taking $f(r)=0$, i.e.
\begin{eqnarray}
  r_{\pm} &=& M+\sqrt{M^2-q^2+\sigma_{r}}, \label{3}
\end{eqnarray}
where $\sigma_{r}$ is parameter of radiation. We will assume only
the cases in which $M^2-q^2+\sigma_{r}\geq 0$ because naked
singularities ($M^2-q^2+\sigma_{r}< 0$) do not occur in nature if
the cosmic conjecture is true. Eq. (\ref{3}) give the location of
event as well as Cauchy horizon of BH, respectively.

Similarly, for Kiselev BH surrounded by dust ($\omega=0$), the
radial coordinated of horizons are
\begin{eqnarray}
  r_{\pm} &=& \frac{2M+\sigma_{d}\pm\sqrt{(2M+\sigma_{d})^2-4q^2}}{2}, \label{5}
\end{eqnarray}
where $\sigma_{d}$ is parameter of dust. We choose the case in which
$(2M+\sigma_{d})^2\geq 4q^2$ because naked singularities
($(2M+\sigma_{d})^2 < 4q^2$) do not occur in nature if the cosmic
conjecture is true \cite{25}. Eq. (\ref{5}) give the location of
event as well as Cauchy horizon of the BH, respectively \cite{11}.

\subsection{Radial Geodesics}

Radial geodesic motion for line element (\ref{1}) in sphereically
symmetric spacetimes is obtained by considering $ds=d\tau$ in Eq.
(\ref{1}), which is \cite{26}
\begin{equation}\label{7}
f(r) \dot{t}^2-f(r)^{-1} \dot{r}^2=1,
\end{equation}
where dot represents the derivative with respect to proper time
$\tau$. Because of the assumption of radial motion, we have
$\dot{\theta}=\dot{\phi}=0$. $E = f(r)\dot{t}$ is well known
conserved energy. By putting this in Eq. (\ref{5}), we have
\begin{equation}\label{8}
\frac{\dot{r}^2}{2}=\frac{E^2-f(r)}{2}.
\end{equation}
For the radial infall of a test particle from rest at position $b$,
we get $E=\sqrt{f(r=b)}$ from Eq. (\ref{8}) \cite{27}. Newtonian
radial acceleration is defined by \cite{28}
\begin{equation}\label{9}
A^{R} = \ddot{r},
\end{equation}
Using Eqs.(\ref{8}) and (\ref{9}), we obtain
\begin{equation}\label{10}
A^{R} = -\frac{f'(r)}{2},
\end{equation}
where prime represents the derivative with respect to $r$ (radial
coordinate). For Kiselev BH surrounded by radiation and dust, we
obtain
\begin{equation}\label{11}
A_r^{R} =-\frac{M}{r^2}+\frac{q^2}{r^3}-\frac{\sigma_{r}}{r^3},
\quad A_d^{R}
=-\frac{M}{r^2}+\frac{q^2}{r^3}-\frac{\sigma_{d}}{r^2}.
\end{equation}
The terms $\frac{q^2}{r^3}-\frac{\sigma_{r}}{r^3}$ and
$\frac{q^2}{r^3}-\frac{\sigma_{d}}{r^2}$ in Eq. (\ref{11})
represents the purely relativistic effect. Eq. (\ref{11}) explain
the "exertion" of Kiselev space-time surrounded by radiation and
dust on a neutral free falling of massive test body. Interestingly,
free fall test particle from rest at $r=b$ (for
$q,\sigma_{r},\sigma_{d}\neq0$) would bounce back at radius
$R^{stop}$. The radius $R^{stop}$ for Kiselev space-time surrounded
by radiation ($R_r^{stop}$) and dust ($R_d^{stop}$) could be found
as
\begin{eqnarray}\label{13}
R_r^{stop}&=&\frac{b(q^2-\sigma_{r})}{2Mb-q^2+\sigma_{r}},
\\\label{13} R_d^{stop}&=&\frac{b~q^2}{(2Mb+b\sigma_{d}-q^2)},
\end{eqnarray}
where $b$ is the initial position starting from rest of test
particle. $R^{stop}$ located inside the Cauchy (internal) horizon.
One can find in the limit $b\rightarrow \infty$,
$R_r^{stop}\rightarrow (q^2-\sigma_{r})/2M$ for radiation and
$R_d^{stop}\rightarrow q^2/(2M-\sigma_{d})$ for dust. The particle
in Kiselev BH surrounded by radiation and dust would emerge in
different asymptotically flat region in maximal analytic extension.
Thus, the particle is physically unstable in maximal analytic
extension beyond the internal (Cauchy) horizons of Kiselev
space-time surrounded by radiation and dust.

\section{Tidal forces in Kiselev space-time on a neutral body in radial free fall}

The equation for the space-like components of the geodesic deviation
vector $\zeta^{\alpha}$ that describes the distance between two
infinitesimally close particles in free fall is given by
\cite{16,17}
\begin{equation}\label{g1}
    \frac{D^2\zeta^{\alpha}}{D \tau^2}-R^{\alpha}_{\beta \gamma
    \delta} v^{\beta} v^{\gamma} \zeta^{\delta} = 0,
\end{equation}
where $v^{\gamma}$ is the unit vector tangent to the geodesic. We
use the tetrad basis for radial free fall reference frames \cite{29}
\begin{eqnarray}
  \hat{e}^{\alpha}_{\hat{0}} &=& \left(\frac{E}{f(r)}, -\sqrt{E^2-f(r)},0,0\right), \\
  \hat{e}^{\alpha}_{\hat{1}} &=& \left( \frac{-\sqrt{E^2-f(r)}}{f(r)},E,0,0\right), \\
  \hat{e}^{\alpha}_{\hat{2}} &=& r^{-1}(0,0,1,0), \\
 \hat{e}^{\alpha}_{\hat{3}} &=& (r \sin(\theta)^{-1}(0,0,0,1),
\end{eqnarray}
where $(x^0,x^1,x^2,x^3)=(t,r,\theta,\phi)$. These unit vectors
satisfy the following orthonomality condition
\begin{equation}\label{g2}
 \hat{e}^{\mu}_{\hat{\alpha}} \hat{e}_{\hat{\gamma}\mu}=\zeta_{\hat{\alpha}
 \hat{\gamma}},
\end{equation}
where $\zeta_{\hat{\alpha} \hat{\gamma}}$ is the Minkowski metric
\cite{16}. We have $\hat{e}^{\alpha}_{\hat{0}}=v^{\alpha}$. The
geodesic deviation vector, also called Jacobi vector, can be written
as
\begin{equation}\label{g3}
    \zeta^{\alpha}=\hat{e}^{\alpha}_{\hat{\gamma}}\zeta^{\hat{\gamma}}.
\end{equation}
Here we note that $\zeta^{\hat{0}}=0$ \cite{16} and
$\hat{e}^{\alpha}_{\hat{\gamma}}$ are all parallelly transported
vectors along the geodesic.

The non-zero independent components of the Riemann tensor in
spherically symmetric space-times are
\begin{eqnarray}
R^{1}_{010} &=& \frac{f(r) f''(r)}{2}, R^{1}_{212} = -\frac{r
f'(r)}{2}, R^{1}_{313} = -\frac{r f'(r)}{2}\sin^2\theta \nonumber \\
R^{2}_{020} &=& \frac{f(r) f'(r)}{2r}, R^{2}_{323} =
\left(1-f(r)\right)\sin^2\theta,R^{3}_{030} = \frac{f(r)
f'(r)}{2r}.\nonumber
\end{eqnarray}

Using above equations in Eq. (\ref{g1}), we find the following
equations for radial free fall tidal forces \cite{21}
\begin{eqnarray}
  \ddot{\zeta}^{\hat{1}}  &=& -\frac{f''}{2}\zeta^{\hat{1}}, \label{g4} \\
  \ddot{\zeta}^{\hat{i}} &=& -\frac{f'}{2r}\zeta^{\hat{i}}, \label{g5}
\end{eqnarray}
where $i = 2, 3$. For aforementioned cases of Kiselev BH,
Eqs.(\ref{g4}) and (\ref{g5}) provided that the tidal forces depend
on the mass and electric charge of a BH as well as radiation and
dust fluid. Tidal forces are identical to Newtonian tidal forces
with the force $-\frac{f'}{2}$ in radial direction can be observed
in Eqs.(\ref{g4}) and (\ref{g5}). Further, we will explore Eqs.
(\ref{g4}) and (\ref{g5}) for Kiselev space-time in detail.

\subsection{Radial tidal forces}

Radial tidal forces vanishes at $r=R^{rtf}_0$ (for radiation) and
$r=R^{rtf}_1$ (for dust) by using Eqs. (\ref{g4}) and (\ref{g5}),
\begin{eqnarray}
  R^{rtf}_0 &=& \frac{3(q^2-\sigma_{r})}{2M}, \label{14} \\
  R^{rtf}_1 &=& \frac{3q^2}{2M+\sigma_{d}}. \label{15}
\end{eqnarray}
The maximum value of radial tidal force is at $R^{rtf}_{0max}$ for
radiation and $R^{rtf}_{1max}$ for dust such that
\begin{eqnarray}
  R^{rtf}_{0max} &=& \frac{2(q^2-\sigma_{r})}{M}, \label{16} \\
  R^{rtf}_{1max} &=& \frac{2q^2}{M+\sigma_{d}}. \label{17}
\end{eqnarray}
The maximum radial stretching for above equations are
\begin{eqnarray}
  \ddot{\zeta}^{1}|_{max} &=& \frac{M^4}{16(q^2-\sigma_{r})^4}, \label{18}\\
  \ddot{\zeta}^{1}|_{max} &=& \frac{(M+\sigma_{d})^4}{16q^6}.  \label{19}
\end{eqnarray}
The radial tidal force using Eqs.(\ref{14}) and (\ref{15}) for
Kiselev BH surrounded by radiation and dust for different choices of
charge is shown in Figures \textbf{1} and \textbf{2}. The local
maximum of radial tidal force for radiation is greater as compared
to the dust near the singularity.
\begin{figure}
  \centering
   \includegraphics[width=8cm]{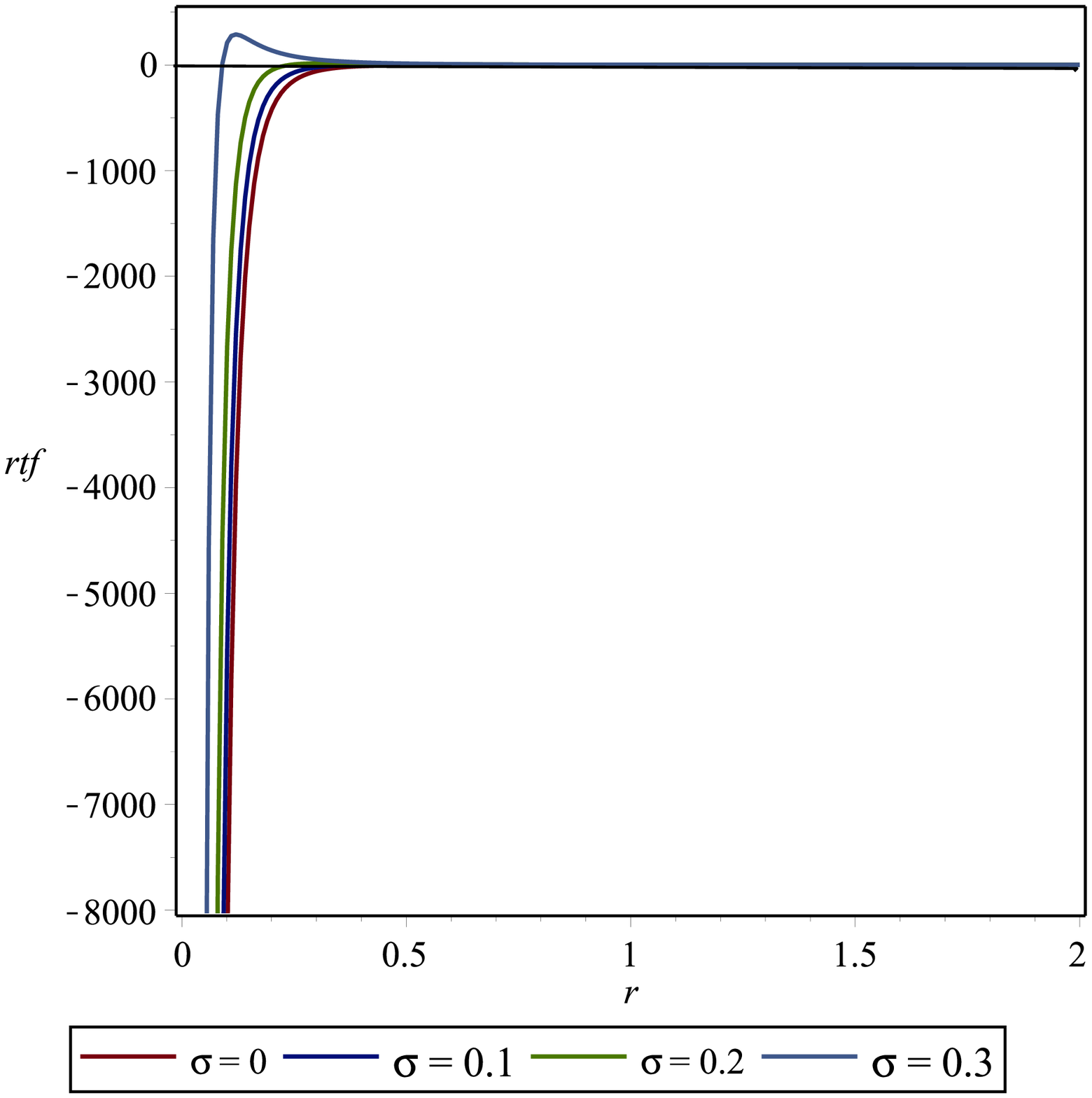}\llap{\raisebox{2.5cm}{\includegraphics[width=4cm]{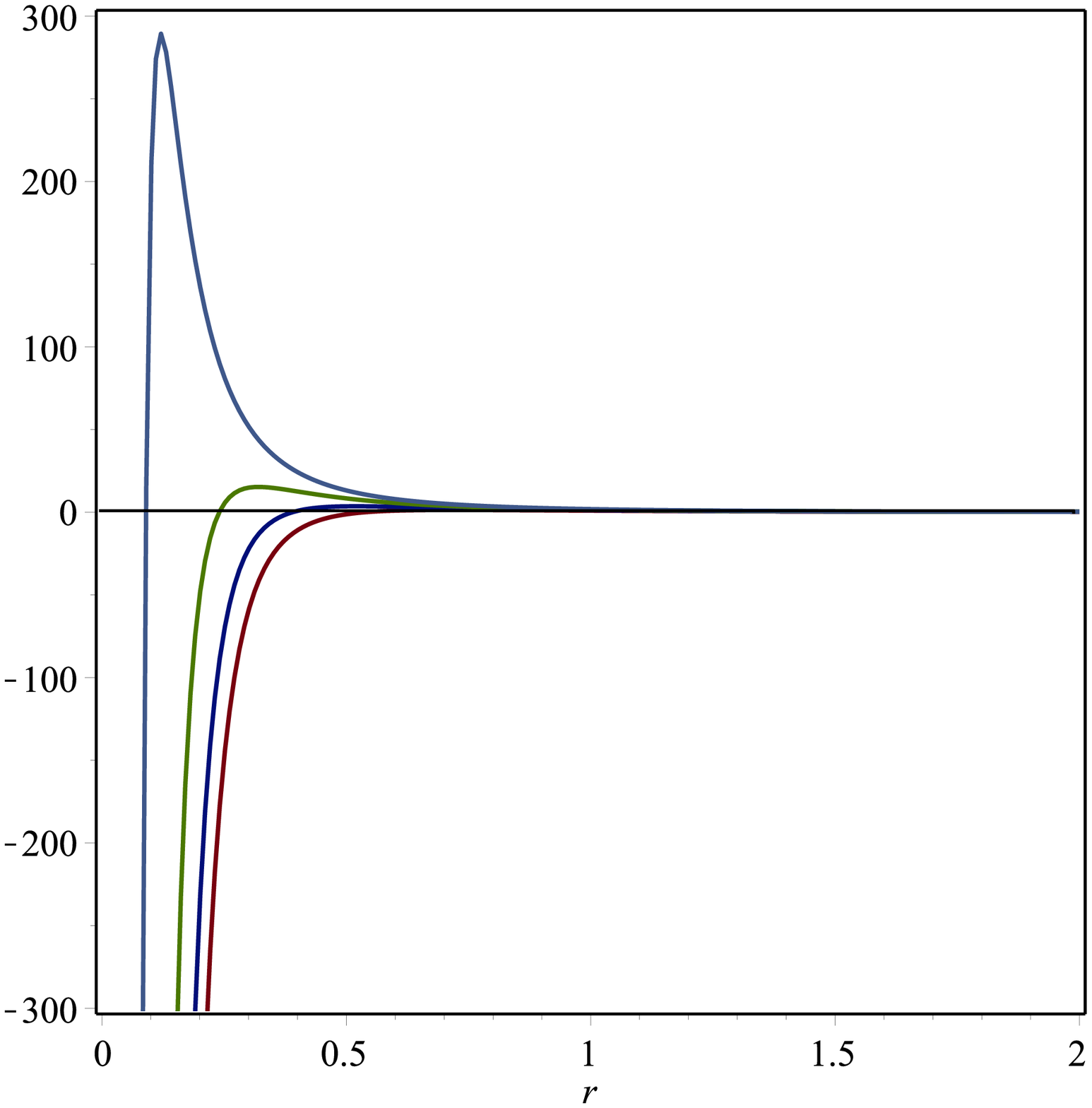}}}\\
  \caption{Radial tidal force for Kiselev BH surrounded by radiation for chosen values radiation parameter, also $q=0.6$ and $b = 100$.}
\end{figure}

\begin{figure}
  \centering
   \includegraphics[width=8cm]{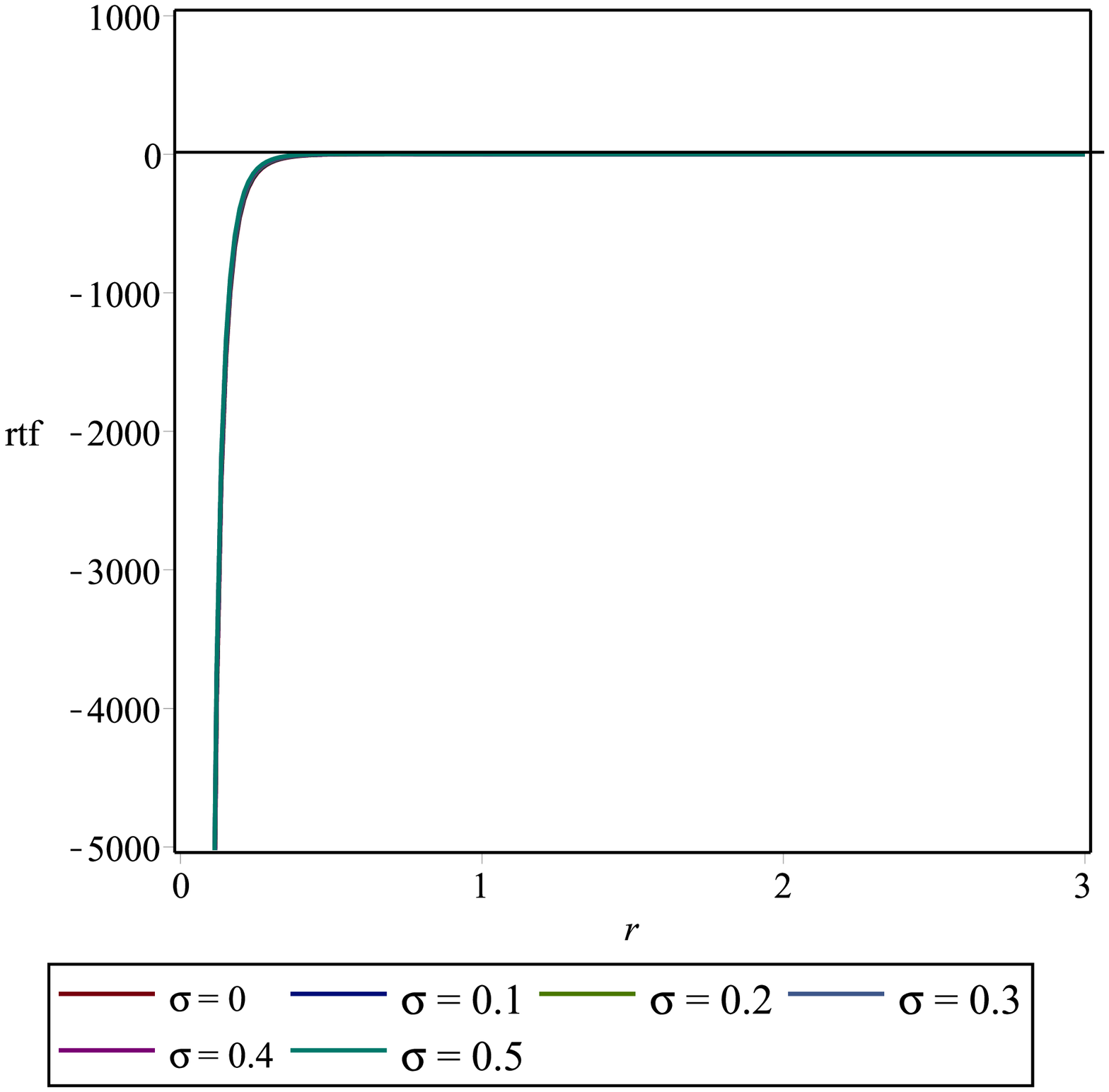}\llap{\raisebox{2.5cm}{\includegraphics[width=4cm]{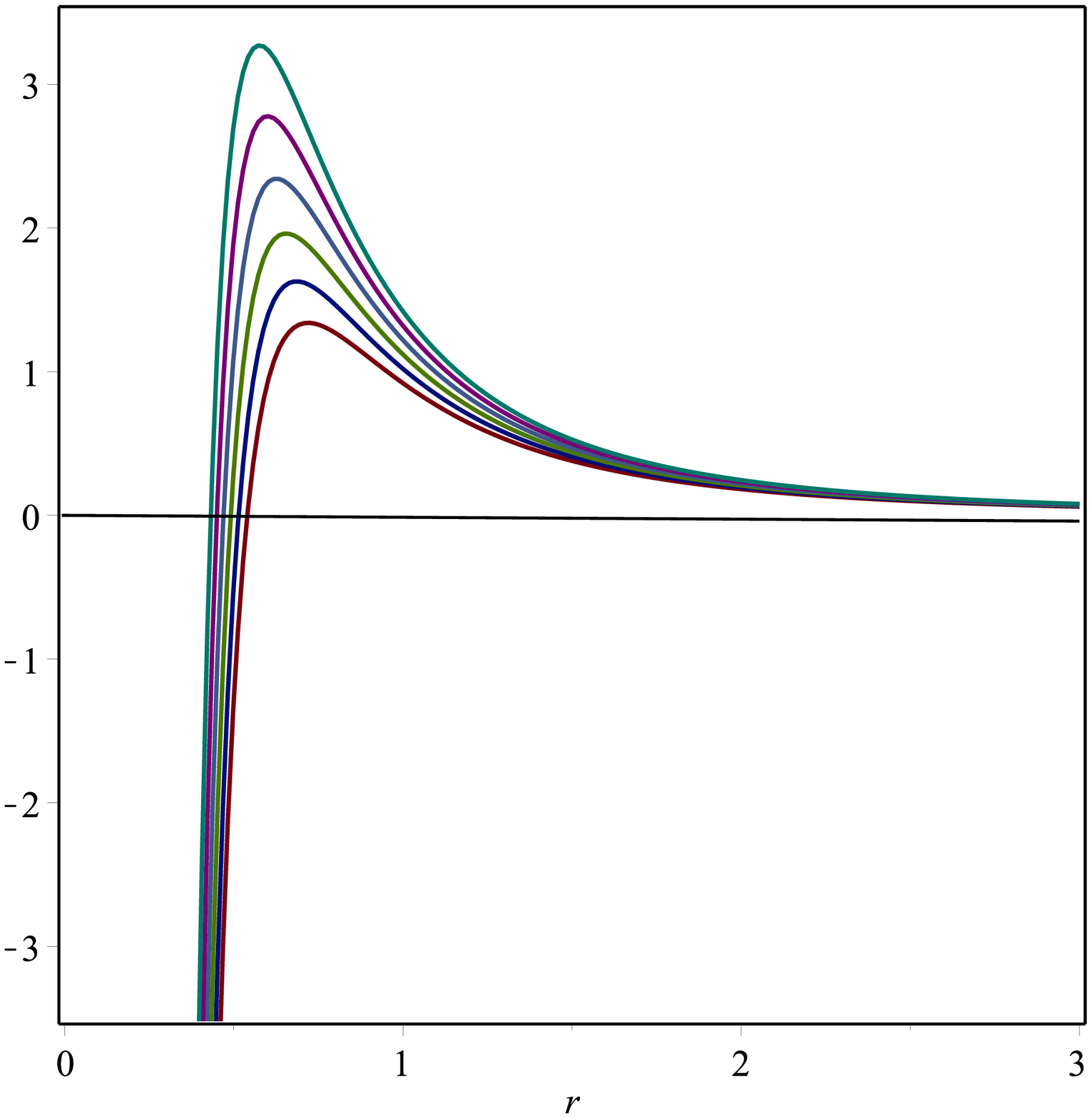}}}\\
  \caption{Radial tidal force for Kiselev BH surrounded by dust for chosen values dust parameter, also $q=0.6$, $M=1$ and $b = 100$.}
\end{figure}

\subsection{Angular tidal forces}
The angular tidal forces vanish at
\begin{eqnarray}
  R^{atf}_0 &=& \frac{q^2-\sigma_r}{M}, \label{20} \\
   R^{atf}_1 &=& \frac{2q^2}{(2M+\sigma_d)}, \label{21}
\end{eqnarray}
by using Eqs. (\ref{2}) and (\ref{g5}) for Kiselev spacetime
surrounded by radiation and dust, respectively. Also one can find
the following conditions from Eqs. (\ref{3}), (\ref{5}), (\ref{20})
and (\ref{21})
\begin{eqnarray}
  r_{-} &\leq& R^{atf}_0 \leq r_{+}, \label{22} \\
  r_{-} &\leq& R^{atf}_1 \leq r_{+}, \label{23}.
\end{eqnarray}
\begin{figure}
  \centering
   \includegraphics[width=8cm]{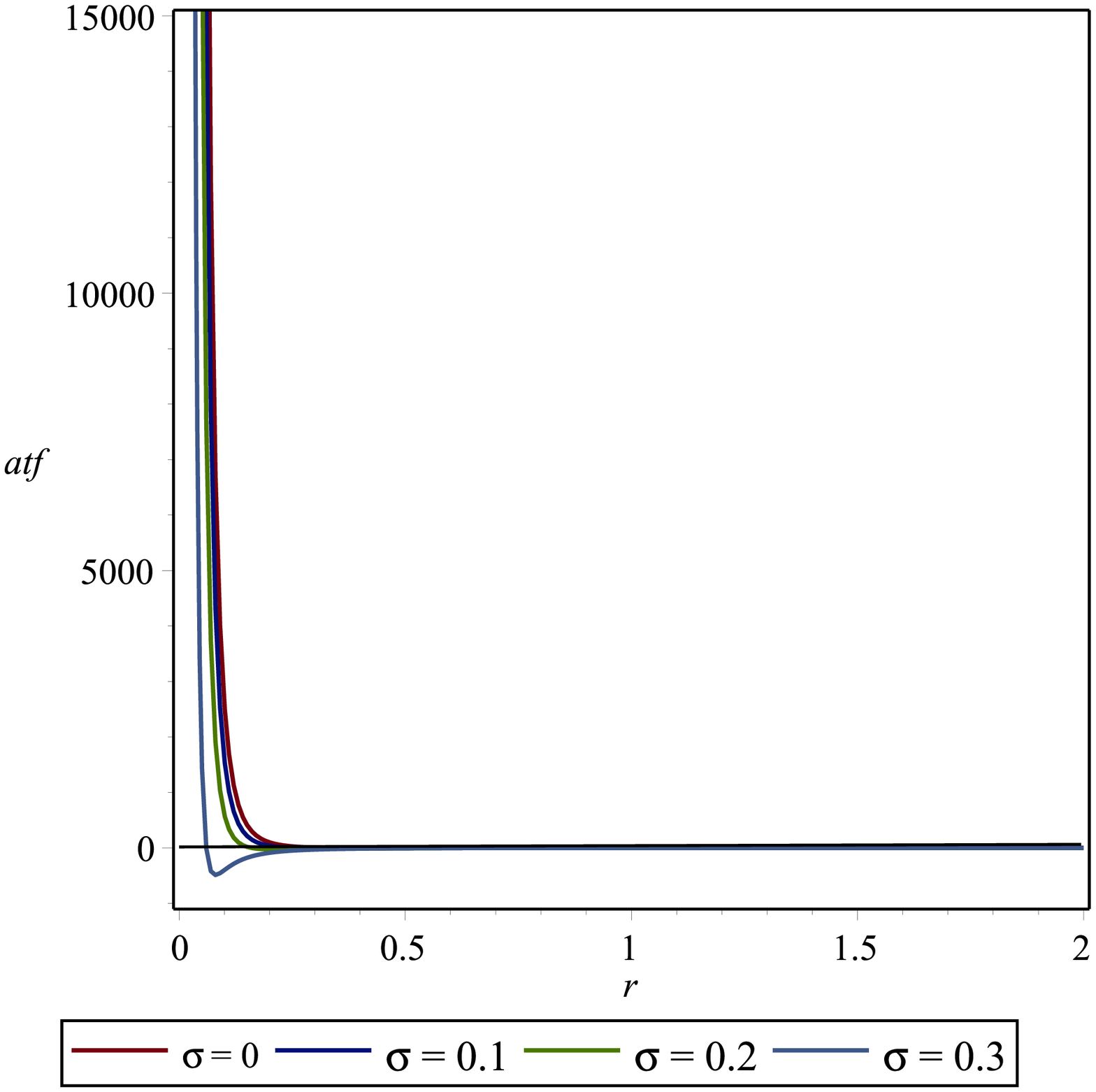}\llap{\raisebox{3.5cm}{\includegraphics[width=4cm]{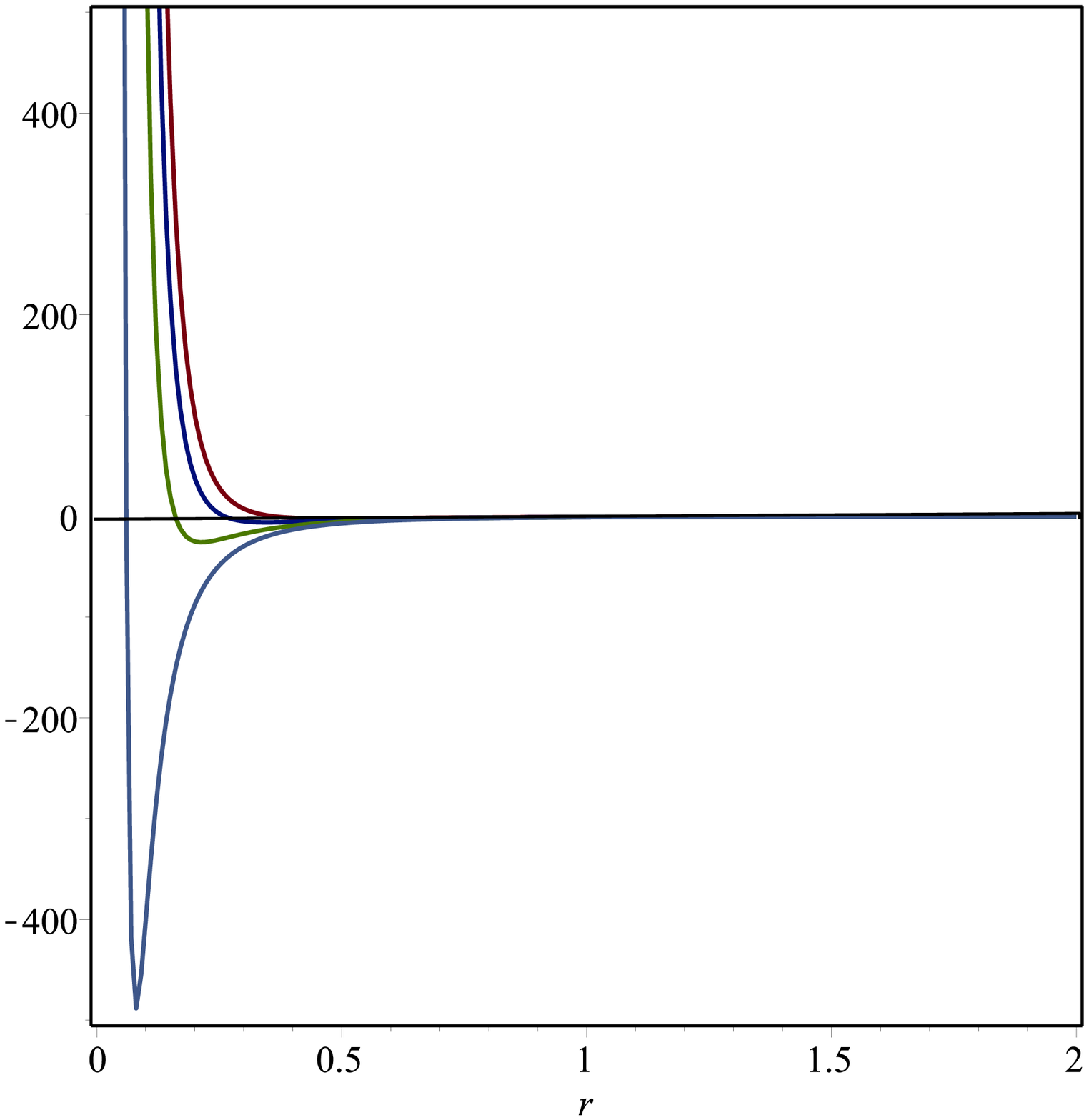}}}\\
  \caption{Angular tidal force for Kiselev BH surrounded by radiation for chosen values radiation parameter, also $q=0.6$, $M=1$ and $b = 100$. }
\end{figure}
\begin{figure}
  \centering
   \includegraphics[width=8cm]{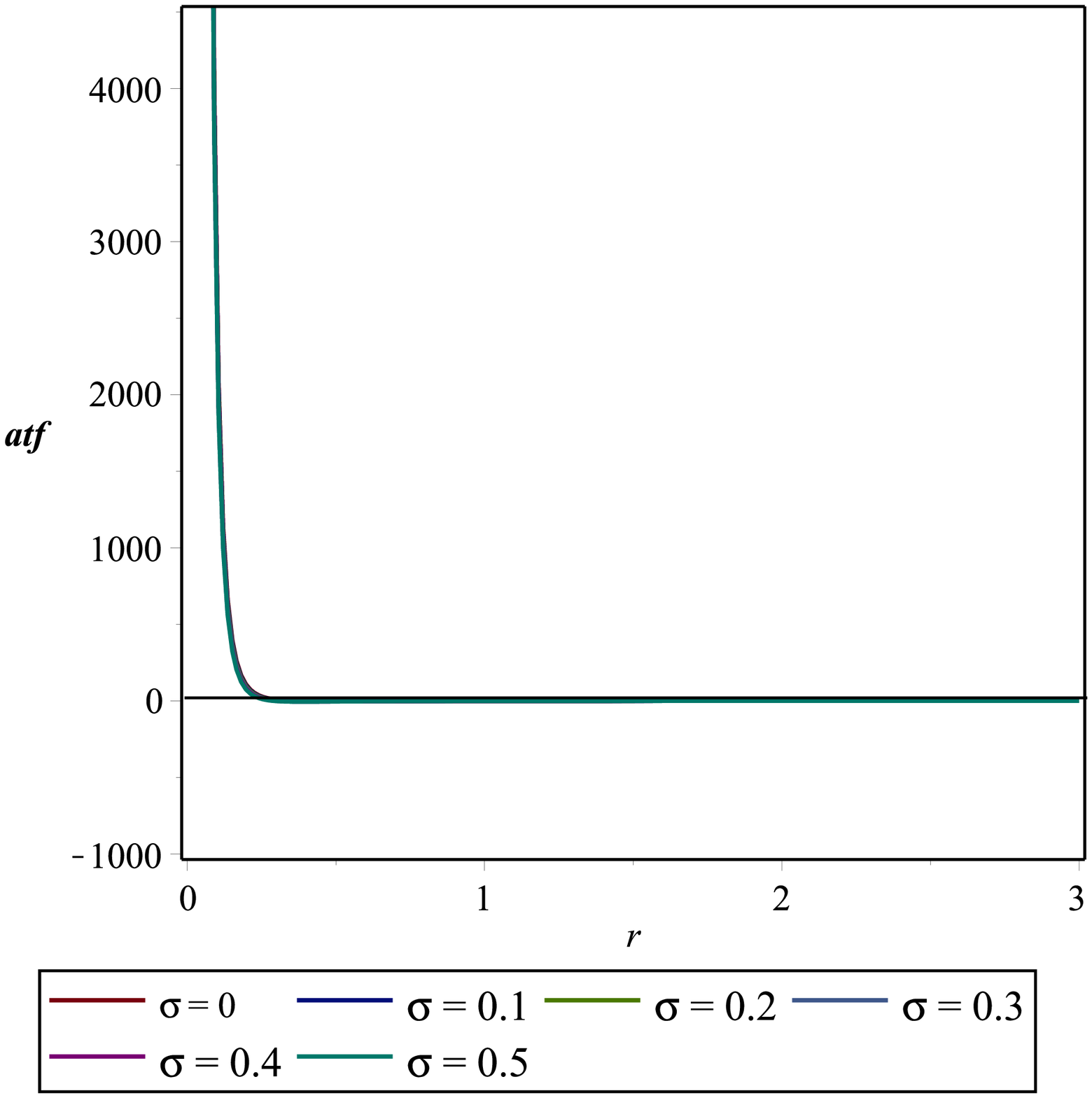}\llap{\raisebox{3.7cm}{\includegraphics[width=4cm]{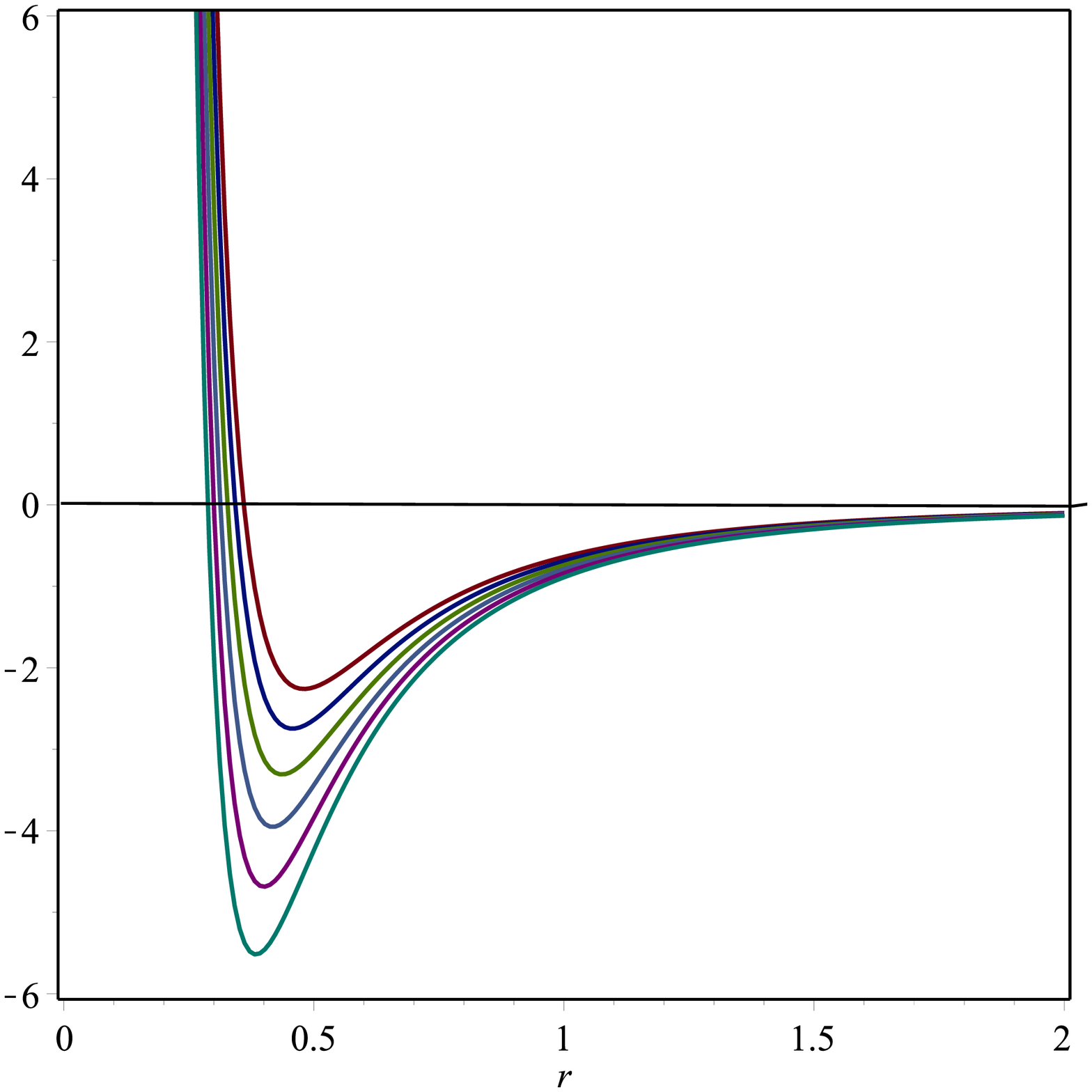}}}\\
  \caption{Angular tidal force for Kiselev BH surrounded by dust for chosen values dust parameter, also $q=0.6$ and $b = 100$.}
\end{figure}
On the basis of these relations, it is pointed out here that the
angular tidal forces becomes zero at some points between the event
horizon and Cauchy horizon. The angular tidal force for different
choices of $q$ for Kiselev BH surrounded by radiation and dust is
given in Figure \textbf{3} and \textbf{4}. The local minimum of
radial tidal force for radiation is greater as compared to the dust
near the singularity.

\section{Solutions of the geodesic equations for Kiselev BH}

We solve the geodesic deviation Eqs.(\ref{g4}) and (\ref{g5}) and
find the geodesic deviation vectors as functions of $r$ for radially
free-falling geodesics. Eqs.(\ref{g4}) and (\ref{g5}) along with
$\frac{dr}{d\tau} = -\sqrt{E^2-f(r)}$ lead to the following
differential equations
\begin{eqnarray}
  (E^2-f(r))\zeta^{\hat{1}\prime\prime}-\frac{f^{\prime}(r)}{2} \zeta^{\hat{1}\prime} +
  \frac{f^{\prime\prime}(r)}{2} \zeta^{\hat{1}}&=& 0, \label{24}\\
  (E^2-f(r))\zeta^{\hat{i}\prime\prime}-\frac{f^{\prime}(r)}{2} \zeta^{\hat{i}\prime} +
  \frac{f^{\prime\prime}(r)}{2} \zeta^{\hat{i}}&=& 0. \label{25}
\end{eqnarray}
The general solution of radial component using Eq. (\ref{g4}) and
(\ref{g5}) is given by
\begin{equation}\label{26}
    \zeta^{\hat{1}}(r)=\sqrt{E^2-f(r)}\left(C_1+C_2\int\frac{dr}{(E^2-f(r))^{3/2}}\right),
\end{equation}
and similarly the angular component
\begin{equation}\label{27}
    \zeta^{\hat{i}}(r)=r\left(C_3+C_4\int\frac{dr}{r^2(E^2-f(r))^{1/2}}\right),
\end{equation}
where $C_1$, $C_2$, $C_3$ and $C_4$ are integration constants
\cite{29}. Since we are considering two cases of Kiselev space-time
i.e. surrounded by radiation and dust. We consider the geodesic
corresponding to a body released from rest at $r=b$. Then the
solution to the geodesic deviation equations about the geodesic for
the case of radiation is given by:
\begin{eqnarray}\label{28}
\zeta^{\hat{1}}(r)&=&\frac{b^3}{M
b-q^2+\sigma_{r}}\dot{\zeta}^{\hat{1}}(b)\sqrt{-\frac{2M}{b}+\frac{q^2}{b^2}
-\frac{\sigma_{r}}{b^2}+\frac{2M}{r}-\frac{q^2}{r^2}+\frac{\sigma_{r}}{r^2}}\\\nonumber
&+&\zeta^{\hat{1}}(b)\left[\left(\frac{3Mb^2
\sqrt{2Mrb(b-r)-(q^2-\sigma_{r})(b^2-r^2)}(M b-q^2+\sigma_{r})
}{(2Mb-q^2+\sigma_{r})^{5/2}b r}\right)\right.\\\nonumber
&\times&\left.
\arctan\left(\frac{(-2Mb+q^2-\sigma_{r})r+Mb^2}{\sqrt{2Mrb(b-r)-(q^2-\sigma_{r})(b^2-r^2)}\sqrt{2Mb-q^2+\sigma_{r}}}\right)\right.\\\nonumber
&+&\left.\frac{1}{(2Mb-q^2+\sigma_{r})^{2}(M b-q^2+\sigma_{r})b
r}\left(M^2(6Mr-3q^2+3\sigma_{r})b^4\right.\right.\\\nonumber
&-&\left.\left.2M^3r^2b^3+10M^2rb^3(q^2-\sigma_{r})
-4Mb^3(q^2-\sigma_{r})^2+\left(5M^2r^2\right.\right.\right.\\\nonumber
&+&\left.\left.\left.5r(q^2-\sigma_{r})M-2(q^2-\sigma_{r})^2\right)(q^2-\sigma_{r})b^2-4Mr^2(q^2-\sigma_{r})^2b\right.\right.\\\nonumber
&+&\left.\left.r^2(q^2-\sigma_{r})^3\right)\right].
\end{eqnarray}
The angular component turns out to be
\begin{equation}\label{29}
     \zeta^{\hat{i}}(r)=\left(\frac{\zeta^{\hat{i}}(b)}{b}+\frac{2b}{\sqrt{q^2-\sigma_{r}}}\dot{\zeta}^{\hat{i}}(b)tan^{-1}
     \sqrt{\frac{(q^2-\sigma_{r})(b-r)}{2Mbr-(q^2-\sigma_{r})(b+r)}}\right)r.
\end{equation}

Similarly, the solution to the geodesic deviation equations about
the geodesic for the case of dust is given by:
\begin{eqnarray}\label{30}
\zeta^{\hat{1}}(r)&=&\frac{2b^3}{2M
b-2q^2+b\sigma_{d}}\dot{\zeta}^{\hat{1}}(b)\sqrt{-\frac{2M}{b}+\frac{q^2}{b^2}
-\frac{\sigma_{d}}{b}+\frac{2M}{r}-\frac{q^2}{r^2}+\frac{\sigma_{d}}{r}}\\\nonumber
&+&\zeta^{\hat{1}}(b)\left[\left(\frac{3b
\sqrt{(b-r)\left((2M+\sigma_{d})b
r-q^2b-q^2r\right)}\left(\left(M+\frac{\sigma_{d}}{2}\right)
b-q^2\right)}{\left(M+\frac{\sigma_{d}}{2}\right)^{-1}\left(2(M
b+\sigma_{d})-q^2\right)^{5/2} r}\right)\right.\\\nonumber
&\times&\left.
tan^{-1}\left(\frac{\left(M+\frac{\sigma_{d}}{2}\right)(b^2-2br)+rq^2}
{\sqrt{(b-r)\left(((2M+\sigma_{d})r-q^2)b-q^2r\right)\left(((2M+\sigma_{d})b-q^2\right)}}\right)\right.\\\nonumber
&+&\left.\frac{1}{\left(2(M b+\sigma_{d})-q^2\right)^{2} b r
\left(\left(M+\frac{\sigma_{d}}{2}\right)
b-q^2\right)}\left(3\left(M+\frac{\sigma_{d}}{2}\right)^2b^4\right.\right.\\\nonumber
&\times&\left.\left.\left(2(M b+\sigma_{d})-q^2\right)+4q^4
b^3\left(M+\frac{\sigma_{d}}{2}\right)-10q^2 r
b^3\left(M+\frac{\sigma_{d}}{2}\right)^2  \right.\right.\\\nonumber
&-&\left. 2 r^2
b^3\left(M+\frac{\sigma_{d}}{2}\right)^3-\left.2q^4+5rq^2
b^2\left(M+\frac{\sigma_{d}}{2}\right)+5r^2q^2
b^2\left(M+\frac{\sigma_{d}}{2}\right)^2\right.\right.\\\nonumber
&-&\left.\left.4r^2q^4 b\left(M+\frac{\sigma_{d}}{2}\right)+q^6
r^2\right)\right].
\end{eqnarray}
For Kiselev BH dust case, angular component becomes
\begin{equation}\label{29}
     \zeta^{\hat{i}}(r)=\left(\frac{\zeta^{\hat{i}}(b)}{b}+\frac{2b^2}{q}\dot{\zeta}^{\hat{i}}(b)tan^{-1}
     \sqrt{\frac{q^2(b-r)}{2Mbr-bq^2-b r \sigma_{d}-q^2 r}}\right)r.
\end{equation}
\begin{figure}
  \centering
   \includegraphics[width=7cm]{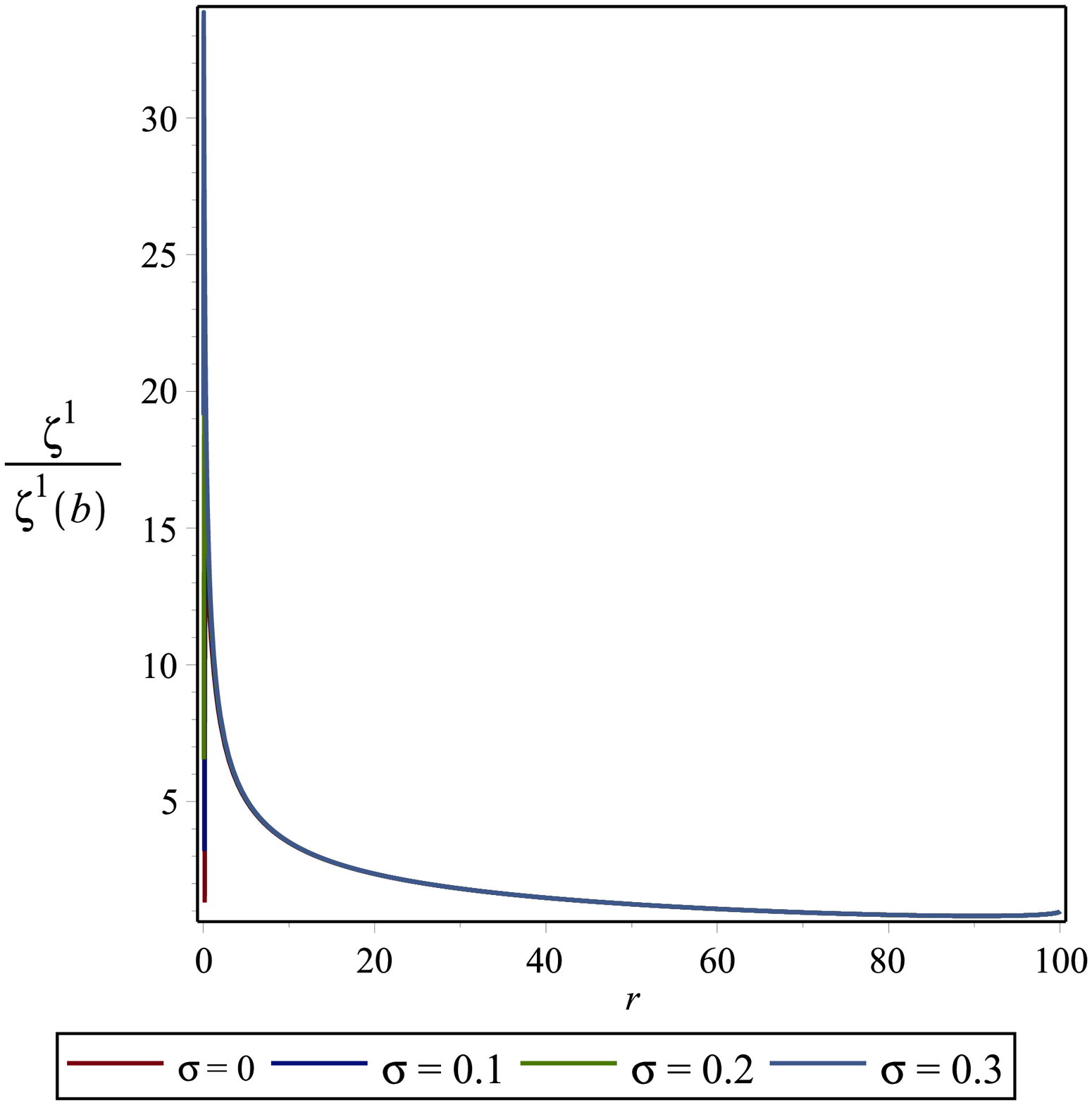}\llap{\raisebox{3cm}{\includegraphics[width=3.5cm]{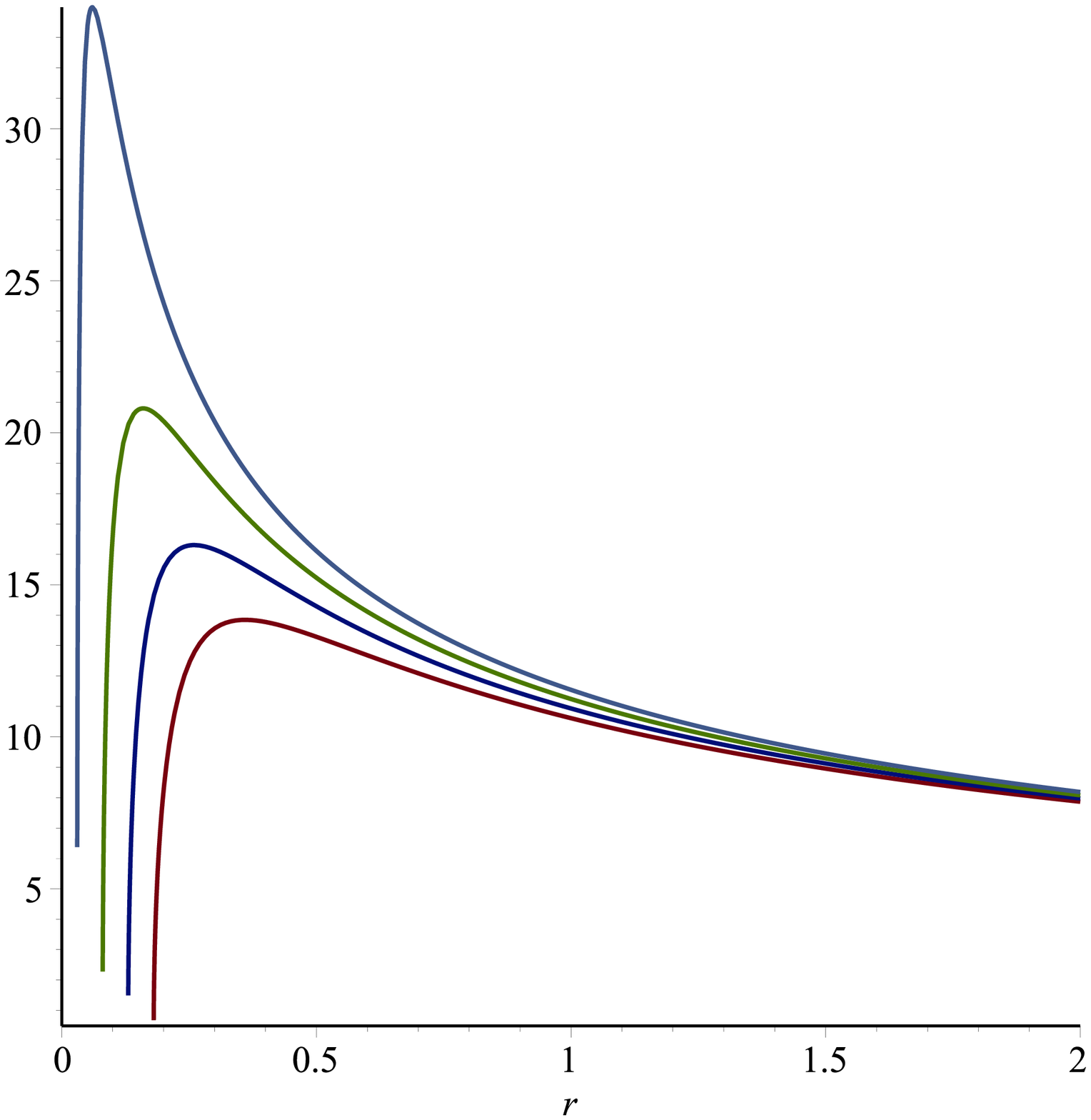}}}\\
  \caption{Radial components of geodesic deviation for Kiselev BH surrounded by radiation with \textbf{IC$1$}
  for different values radiation parameter with $M=1$, $q=0.6$ and $b = 100$.}
\end{figure}
\begin{figure}
  \centering
   \includegraphics[width=7cm]{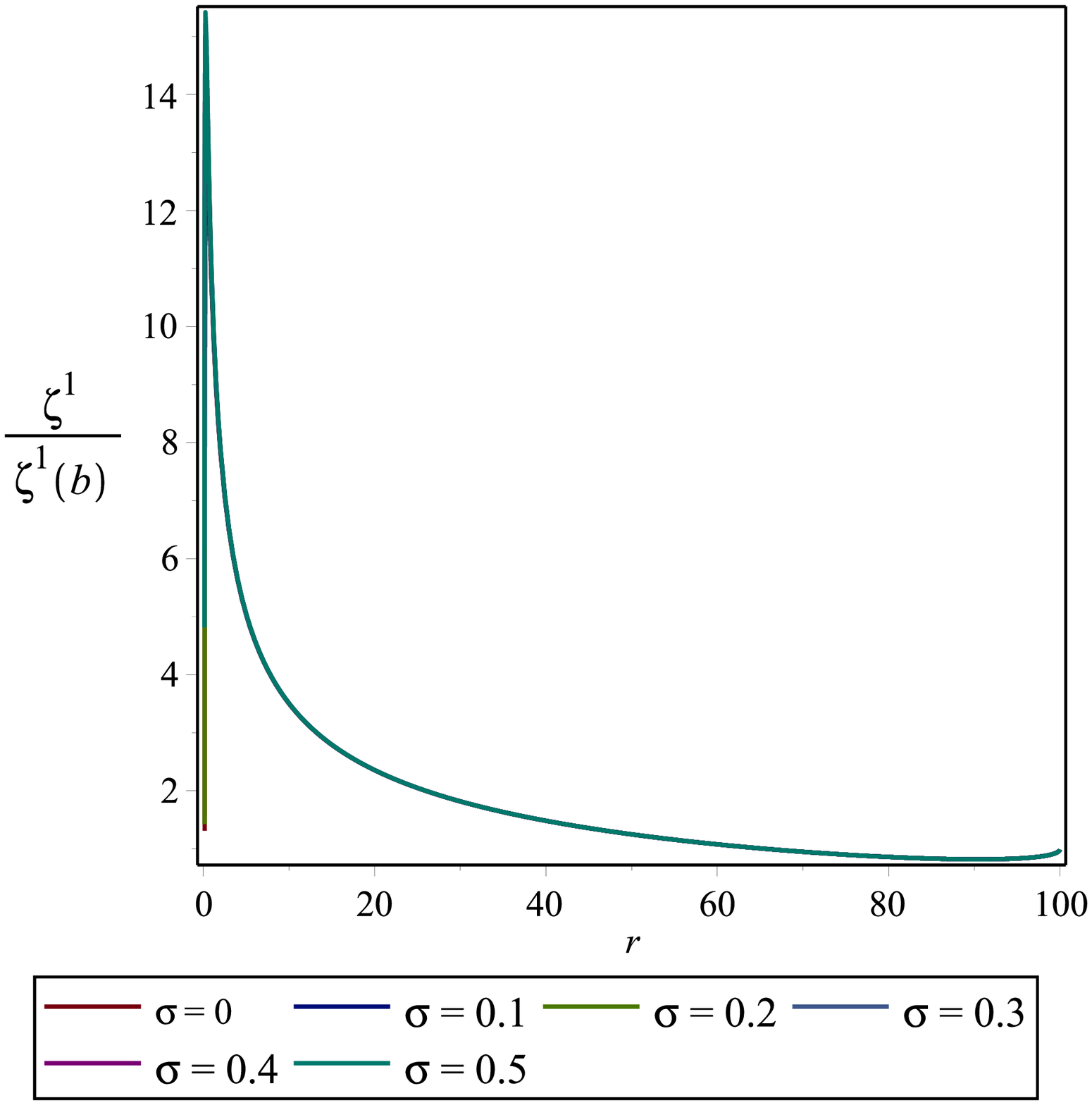}\llap{\raisebox{3cm}{\includegraphics[width=3.5cm]{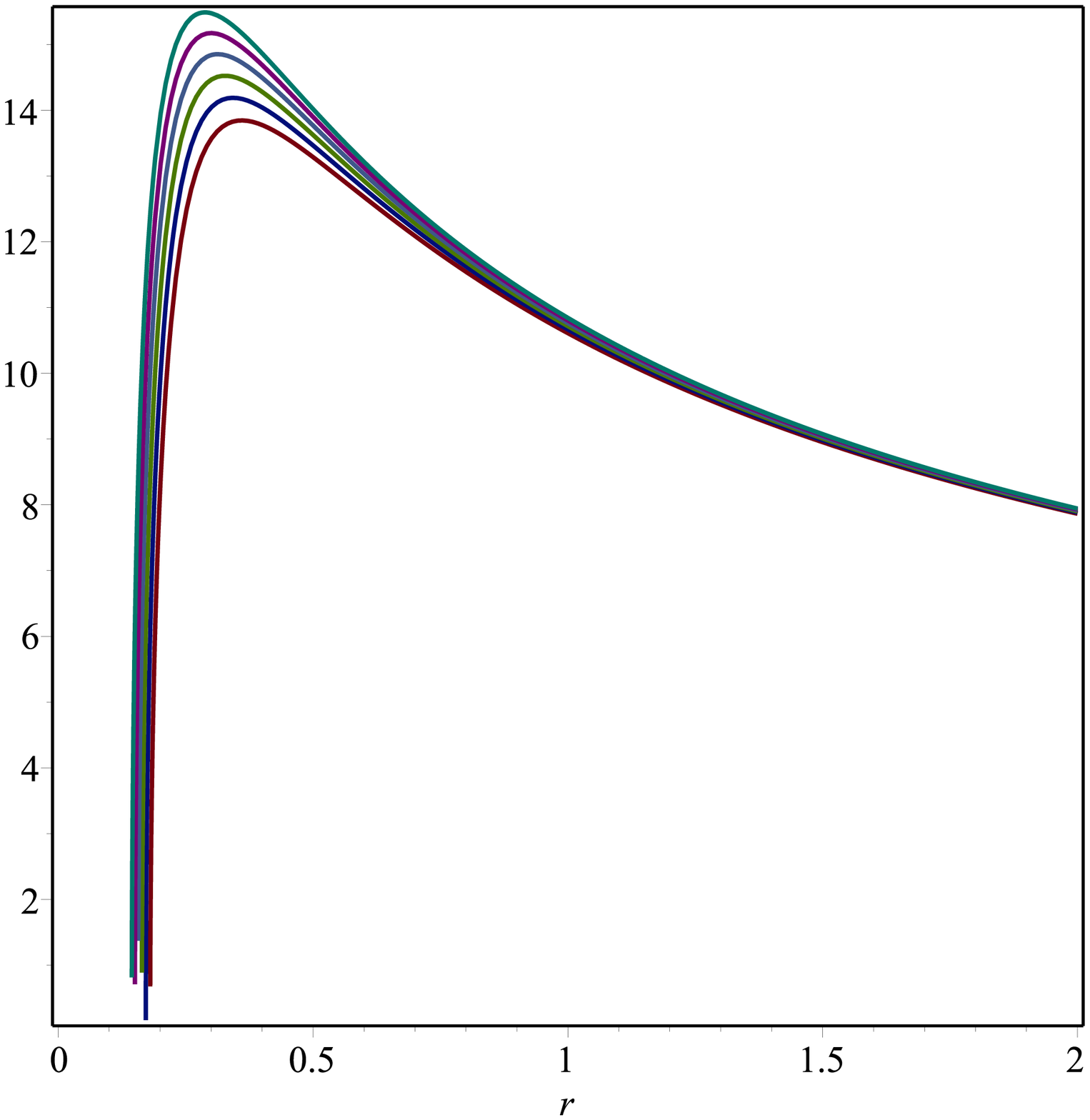}}}\\
  \caption{Radial components of geodesic deviation for Kiselev BH surrounded by dust with \textbf{IC$1$}
   for different values dust parameter with $M=1$, $q=0.6$ and $b = 100$.}
\end{figure}
Here, $\zeta^{\hat{1}}(b)$ and $\zeta^{\hat{i}}(b)$ are the radial
and angular components of the initial geodesic deviation vector at
$r=b$ and $\dot{\zeta}^{\hat{1}}(b)$ and $\dot{\zeta}^{\hat{i}}(b)$
are the corresponding derivatives with respect to the proper time
$\tau$. Figure \textbf{5}-\textbf{10} represents the radial and
angular components of the geodesic deviation vector of a body
in-falling from rest at $r =b$ towards BH for different choices of
the radiation and dust parameters. We choose the initial condition
(IC$1$) $\zeta^{\hat{1}}(b)
> 0$, $\dot{\zeta}^{\hat{1}}(b)=0$ and $\zeta^{\hat{i}}(b) > 0$,
$\dot{\zeta}^{\hat{i}}(b)=0$. It represents the releasing a body at
rest consisting of dust with no internal motion. On the other hand,
we choose the initial condition (IC$2$) $\zeta^{\hat{1}}(b) = 0$,
$\dot{\zeta}^{\hat{1}}(b)>0$ and $\zeta^{\hat{i}}(b) = 0$,
$\dot{\zeta}^{\hat{i}}(b)>0$. It corresponds to letting such a body
explode from a point at $r=b$. The behavior of the geodesic
deviation vector is almost identical for different values of
radiation and dust parameter until \emph{r} becomes of the same
order as the horizon radius.

In Figure \textbf{5}, the radial tidal force for Kiselev BH
surrounded by radiation depends upon the radiation parameter with
\textbf{IC$1$} i.e. it attains the highest maximum value at
$\sigma_r = 0.3$ while as the radiation parameter decrease the
maximum value of radial tidal force also decreases. Also the maximum
value is shifting towards BH as the radiation parameter increases.
For $\sigma_r
> 0.3$, it becomes
unphysical. Figure \textbf{6} represents the radial tidal force for
Kiselev BH surrounded by dust for different choices of dust
parameter with \textbf{IC$1$}. It is clear from figure the maximum
value of radial tidal force is increasing as the value of dust
parameter increases as well as it shifting towards the BH. It is
concluded that the maxima of radial tidal force depends upon
radiation and dust parameter, it increases for large values of
$\sigma_r$ and $\sigma_d$. Figures \textbf{7} and \textbf{8}
represent the radial tidal forces for Kiselev BH surrounded by
radiation and dust parameter for its different choices with
\textbf{IC$2$}. In Figure \textbf{7}, maxima of radial tidal force
is increasing for higher values of radiation parameter and attains
the maximum value at $\sigma_r=0.3$, and also it is shifted towards
the BH as the radiation parameter increases. But in Figure
\textbf{8}, maximum value of radial tidal force is same for all
chosen values of dust parameter. Although, the radial tidal force is
shifting towards the BH as the dust parameter increases.

The angular tidal force for Kiselev BH surrounded by radiation and
dust parameter with \textbf{IC$1$} is similar as explained in
\cite{29}, as it does not depend upon radiation and dust parameter.
However, the angular tidal force with \textbf{IC$2$} are discussed
in Figure \textbf{9} and \textbf{10} for Kiselev BH surrounded by
radiation and dust respectively. It can be seen in Figure \textbf{9}
that the angular tidal force is increasing from $r=100$ and attains
maximum value at $r=50$ then start decreasing, reflecting the
compressing nature of angular component. At some point near the BH,
angular tidal force start increasing as shown in Figure \textbf{9}.
Also, it is shifting towards the singularity as radiation parameter
increases. In Figure \textbf{10}, the angular tidal force have the
maximum value for $\sigma_d = 0$, while it decreases for large
values of dust parameter.

\begin{figure}
  \centering
   \includegraphics[width=8cm]{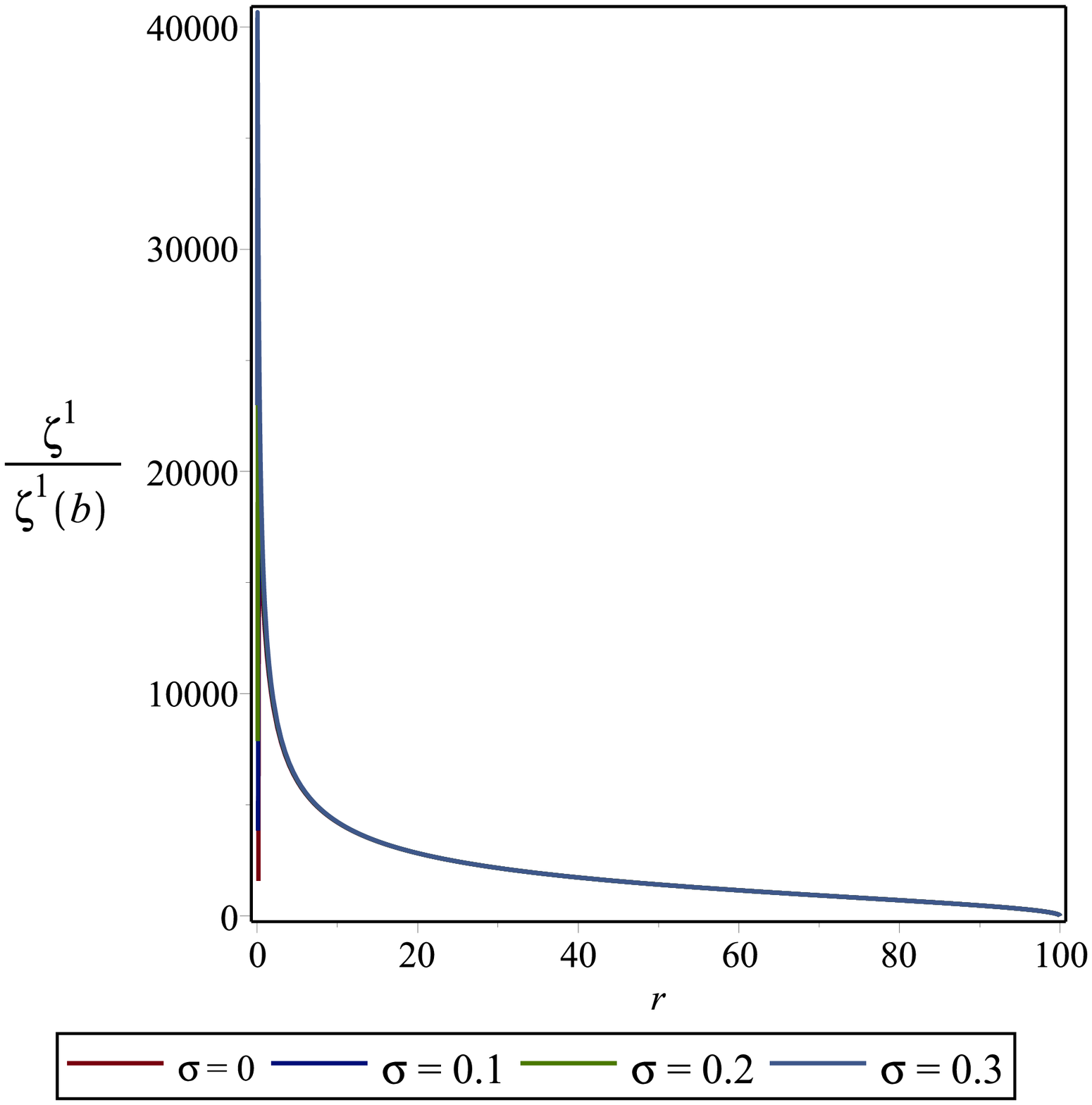}\llap{\raisebox{3.5cm}{\includegraphics[width=4cm]{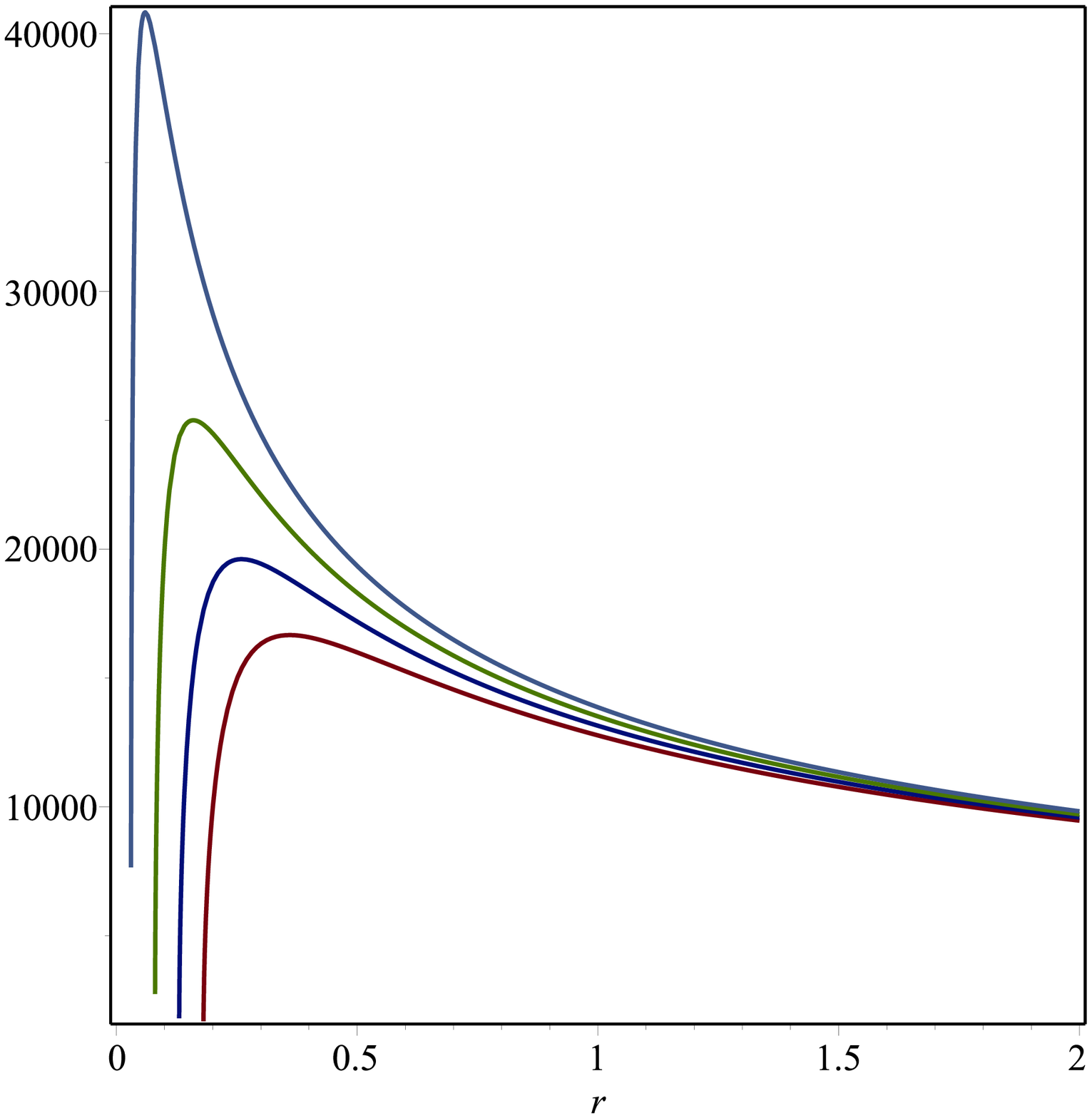}}}\\
  \caption{Radial components of geodesic deviation for Kiselev BH surrounded by radiation with \textbf{IC$2$}
   for different values radiation parameter with $M=1$, $q=0.6$ and $b = 100$.}
\end{figure}

\begin{figure}
  \centering
   \includegraphics[width=8cm]{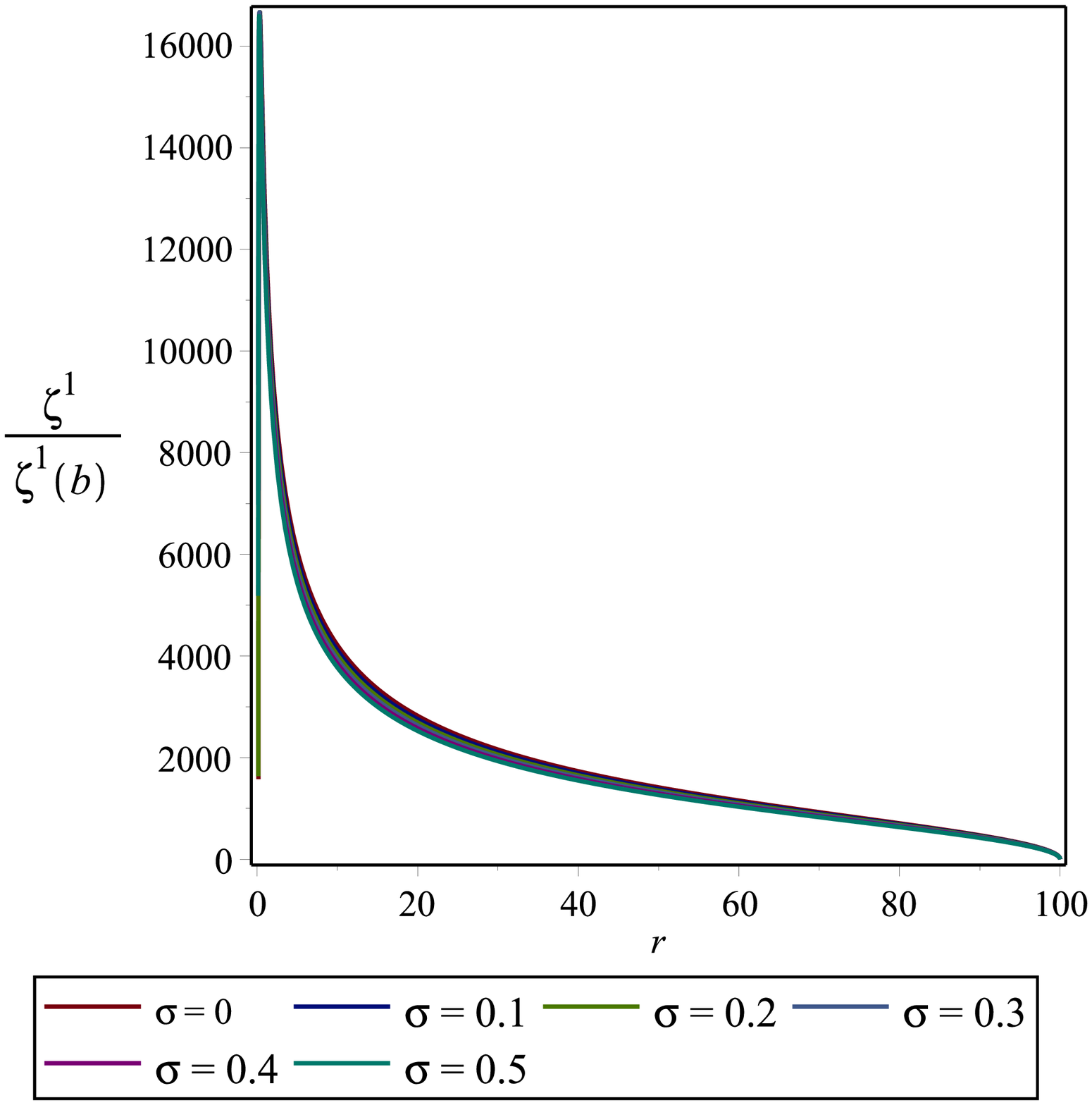}\llap{\raisebox{3.7cm}{\includegraphics[width=4cm]{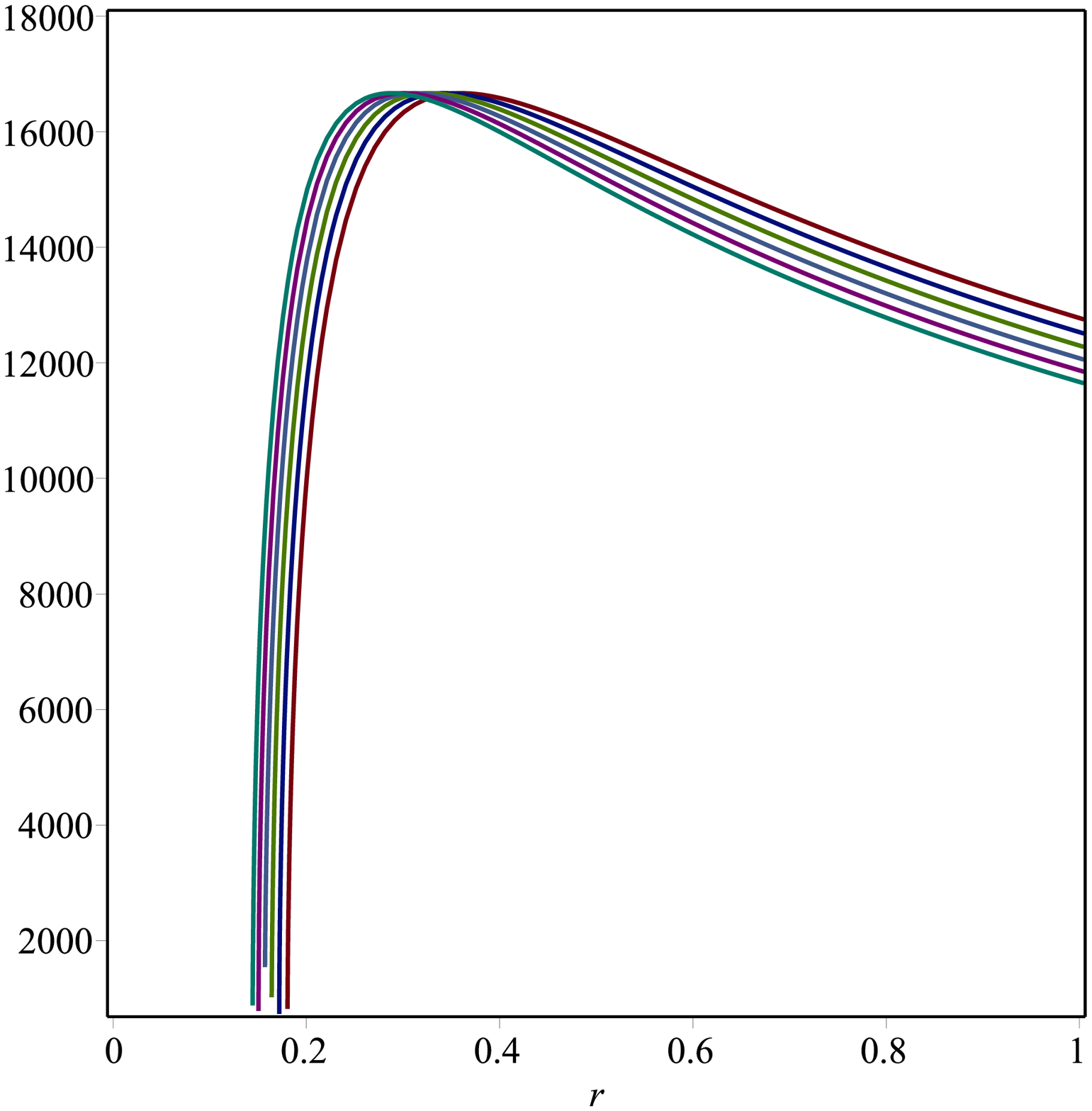}}}\\
  \caption{Radial components of geodesic deviation for Kiselev BH surrounded by dust with \textbf{IC$2$}
   for different values dust parameter with $M=1$, $q=0.6$ and $b = 100$.}
\end{figure}

\begin{figure}
  \centering
   \includegraphics[width=8cm]{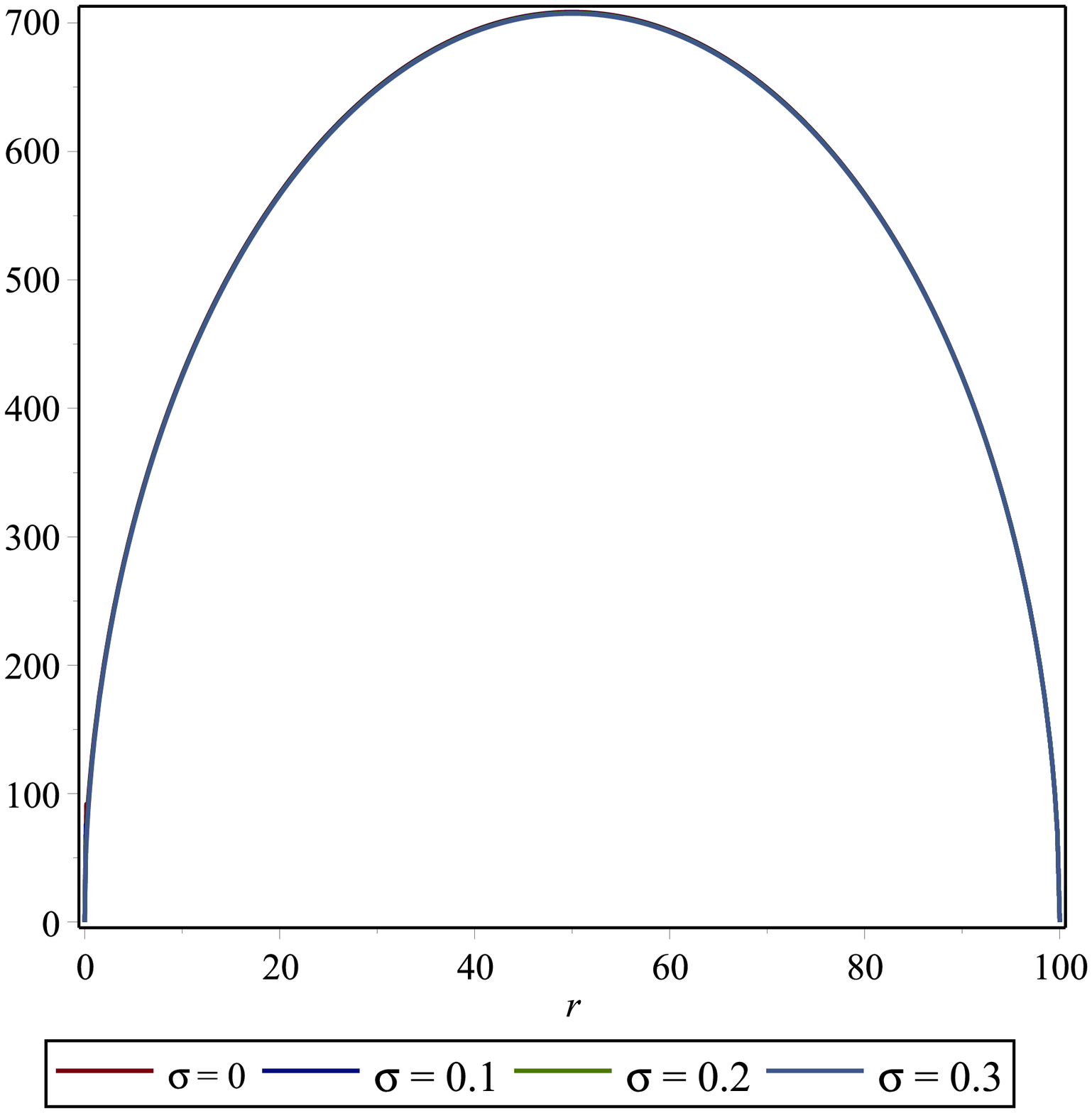}\llap{\makebox[\wd1][l]{\raisebox{3.7cm}{\includegraphics[width=3cm]{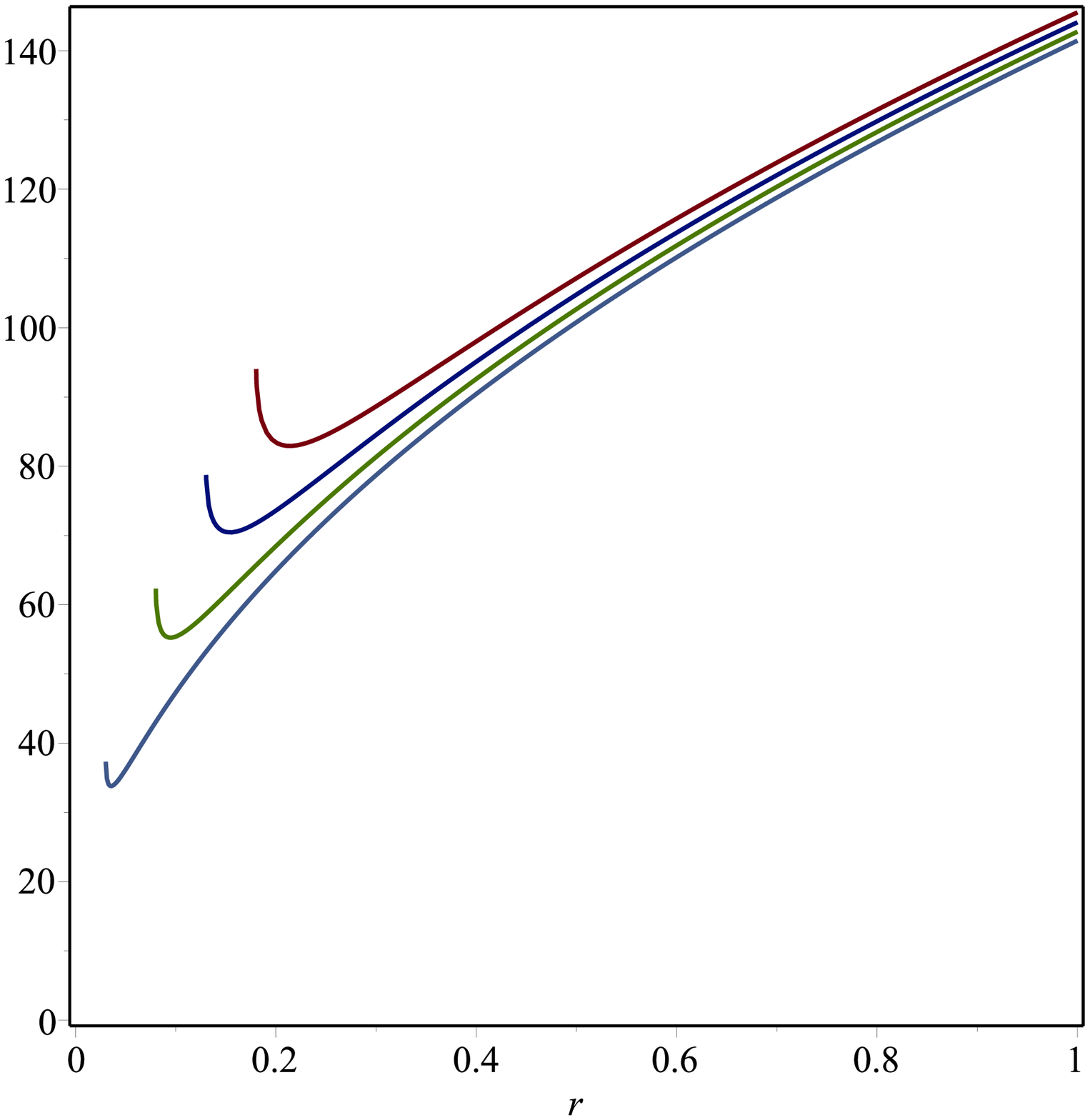}}}}\\
  \caption{Angular components of geodesic deviation for Kiselev BH surrounded by radiation with IC$2$
  for different values radiation parameter with $M=1$, $q=0.6$ and $b = 100$.}
\end{figure}

\begin{figure}
  \centering
   \includegraphics[width=7cm]{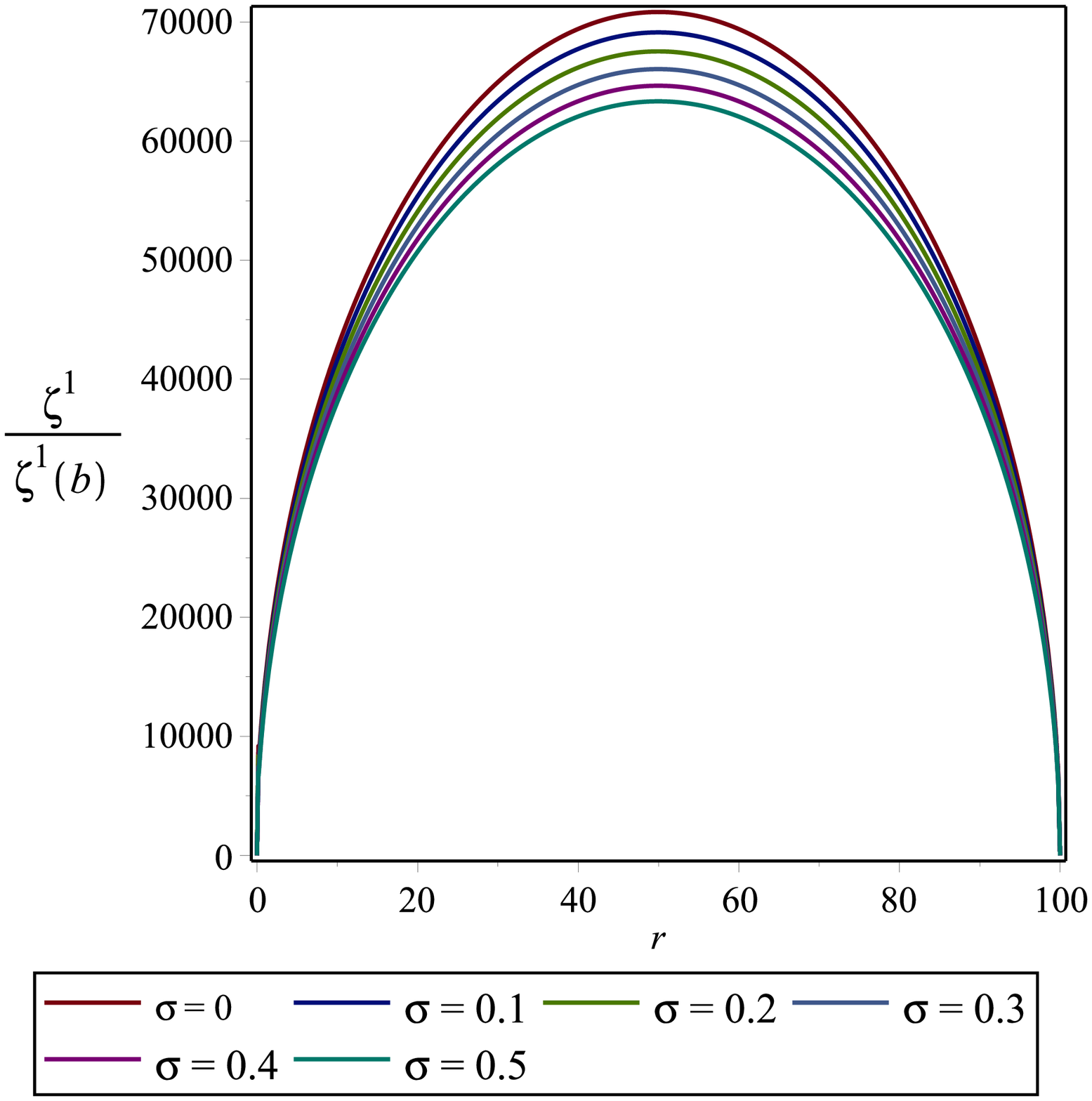}\llap{\makebox[\wd1][l]{\raisebox{3cm}{\includegraphics[width=3cm]{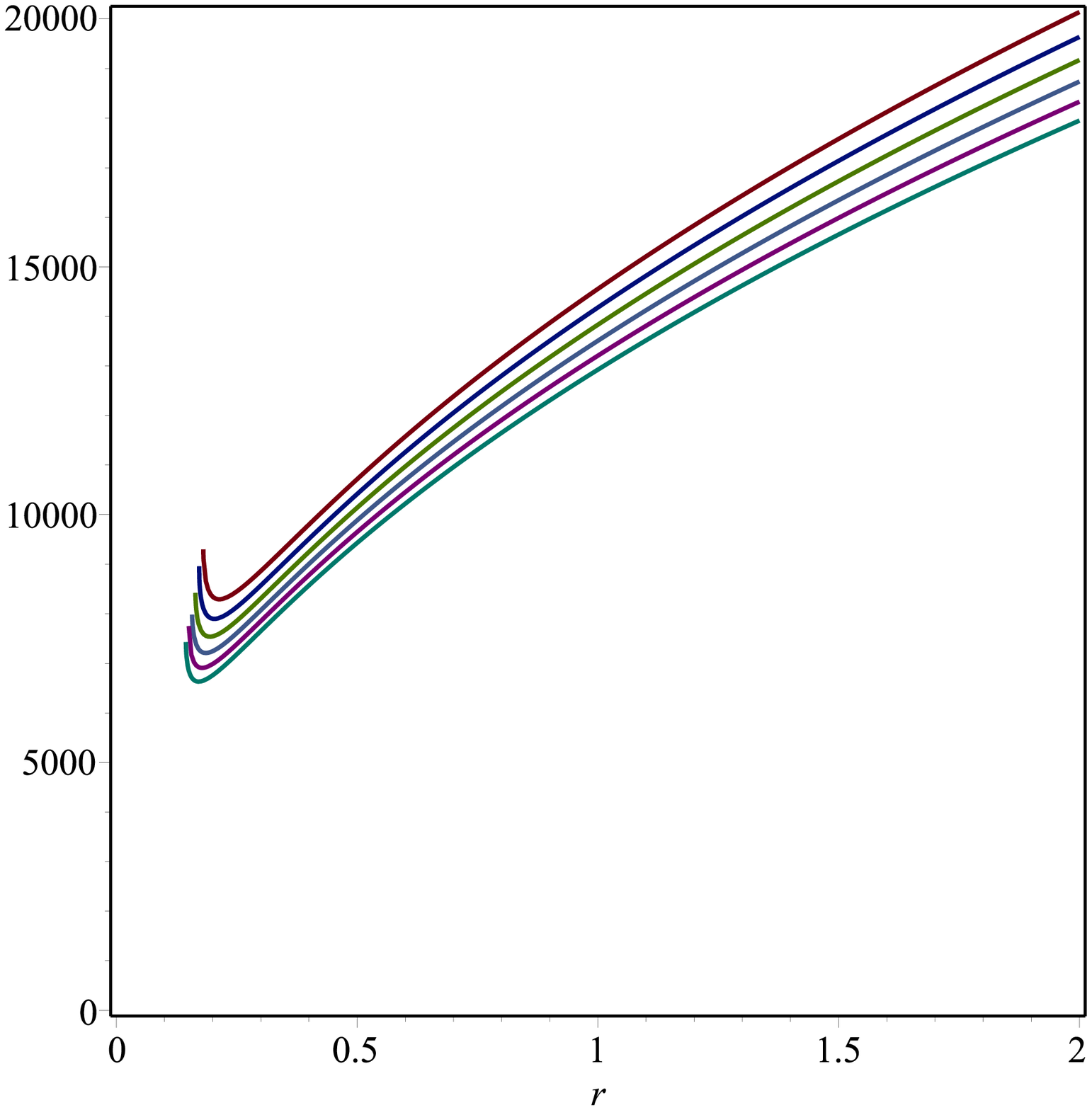}}}}\\
  \caption{Angular components of geodesic deviation for Kiselev BH surrounded by dust with IC$2$
  for different values dust parameter with $M=1$, $q=0.6$ and $b = 100$.}
\end{figure}

\begin{table}
\begin{center}
\begin{tabular}{|c|c|c|c|c|c|c|}
\hline
$\sigma_d$ & $r_+$ & $rtf$ & $atf$ & $Figure~6$ & $Figure~8$ & $Figure~10$ \\
\hline
0 & 1.8 & 0.2400 & -0.1371 & 8.2543 & 9935.44 & 19160.04 \\
0.1 & 1.9116 & 0.2197 & -.1233 & 8.0509 & 9455.87 & 19221.57 \\
0.2 & 2.0219 & 0.2015 & -.1115 & 7.8615 & 9020.038 & 19271.458 \\
0.3 & 2.1310 & 0.1852 & -.1013 & 7.6845 & 8622.22 & 19312.18 \\
0.4 & 2.2392 & 0.1707 & -0.0925 & 7.5186 & 8257.65 & 19345.57 \\
0.5 & 2.3465 & 0.1578 & -0.0848 & 7.3627 & 7922.33 & 19373.02\\
\hline
\end{tabular}
\end{center}
\caption{Location of event horizon for chosen values of dust
parameter.} \label{01}
\end{table}

\begin{table}
\begin{center}
\begin{tabular}{|c|c|c|c|c|c|c|}
\hline
$\sigma_d$ & $r_-$ & $rtf$ & $atf$ & $Figure~6$ & $Figure~8$ & $Figure~10$ \\
\hline
0 & 0.2 & -425.00 & 100.00 & 8.2542 & 9935.44 & 8343.08 \\
0.1 & 0.1883 & -544.31 & 129.03 & 8.0508 & 9455.87 & 7966.709 \\
0.2 & 0.1780 & -684.94 & 163.34 & 7.8615 & 9020.03 & 7625.47 \\
0.3 & 0.1689 & -849.08 & 203.51 & 7.6845 & 8622.22 & 7314.24 \\
0.4 & 0.1607 & -1039.06 & 250.10 & 7.5186 & 8257.65 & 7028.92 \\
0.5 & 0.1534 & -1257.28 & 303.70 & 7.3627 & 7922.33 & 6766.19\\
\hline
\end{tabular}
\end{center}
\caption{Location of Cauchy horizon for chosen values of dust
parameter.} \label{01}
\end{table}

\begin{table}
\begin{center}
\begin{tabular}{|c|c|c|c|c|c|c|}
\hline
$\sigma_r$ & $r_+$ & $rtf$ & $atf$ & $Figure~5$ & $Figure~7$ & $Figure~9$ \\
\hline
0 & 1.8 & 0.24 & -.1371 & 8.2542 & 9935.44 & 191.60 \\
0.1 & 1.8602 & 0.2455 & -.1336 & 8.2522 & 9925.43 & 193.59 \\
0.2 & 1.9165 & 0.2485 & -.1302 & 8.2501 & 9915.44 & 195.40 \\
0.3 & 1.9695 & 0.2498 & -.1269 & 8.2480 & 9905.47 & 197.05 \\
\hline
\end{tabular}
\end{center}
\caption{Location of event horizon for chosen values of radiation
parameter.} \label{01}
\end{table}

\begin{table}
\begin{center}
\begin{tabular}{|c|c|c|c|c|c|c|}
\hline
$\sigma_r$ & $r_-$ & $rtf$ & $atf$ & $Figure~5$ & $Figure~7$ & $Figure~9$ \\
\hline
0 & 0.2 & -425.00 & 100.00 & 8.2542 & 9935.44 & 83.4308 \\
0.1 & .1397 & -1311.44 & 315.06 & 8.2521 & 9925.43 & 71.5674 \\
0.2 & 0.08348 & -6443.99 & 1575.13 & 8.2500 & 9915.44 & 57.0569 \\
0.3 & 0.0304 & -138248.39 & 34292.71 & 8.2479 & 9905.47 & 36.0236 \\
\hline
\end{tabular}
\end{center}
\caption{Location of Cauchy horizon for chosen values of radiation
parameter.} \label{01}
\end{table}
Table \textbf{1} - \textbf{4} represent the location of event
horizon and Cauchy horizon for chosen values of radiation and dust
parameter. Here, we find the location of event and Cauchy horizons
of \textbf{rtf} (Figures \textbf{1} and \textbf{2}), \textbf{atf}
(Figures \textbf{3} and \textbf{4}), radial tidal force with
\textbf{IC$1$} (Figures \textbf{5}, \textbf{6}) and \textbf{IC$2$}
(Figures \textbf{7}, \textbf{8}), respectively and angular tidal
force with \textbf{IC$2$} (Figures \textbf{9} and \textbf{10}) for
chosen values of radiation and dust parameter, respectively. It is
noted that the radial tidal force (in Figures \textbf{1} and
\textbf{2}) and angular tidal force (in Figures \textbf{3} and
\textbf{4}) change their sign between event and Cauchy horizons.
Event horizon is increasing (away from singularity) and Cauchy
horizon is decreasing (shifting towards singularity) as the
radiation and dust parameter increase.

\section{Conclusion}

We investigated the tidal forces of Kiselev BHs by assuming its two
special cases, i.e. Kiselev BH surrounded by radiation and dust. We
have observed that the radial tidal forces can change its behavior
from stretching to compressing for specific choices of radiation and
dust parameters and angular tidal forces can only be zero between
event and Cauchy horizons of BH. It is also mention here that the
radial and angular tidal forces possesses an ability to change their
sign between event and Cauchy horizons. Event horizon is increasing
(away from singularity) and Cauchy horizon is decreasing (shifting
towards singularity) as the radiation and dust parameter increase.
Furthermore, the geodesic deviation equations can be solved
analytically about radially free-falling geodesic for Kiselev BH
\cite{29}. The behavior of geodesic deviation vector for such a
geodesic under the influence of tidal forces are examined. We choose
the initial condition \textbf{IC$1$} which represents the releasing
a body at rest consisting of dust with no internal motion and
\textbf{IC$2$} corresponds to letting such a body explode from a
point at $r=b$. It is pointed out here that the radial tidal forces
for Kiselev BH surrounded by radiation attains the maximum value at
$\sigma_r = 0.3$ as shown in Figure \textbf{5} and it becomes
unphysical for $\sigma_r
> 0.3$. Moreover, the maxima of radial tidal forces for Kiselev BH
surrounded by dust is increasing as well as shifting towards the BH
as the dust parameter increasing. These are the agreement with
\cite{29}. Hence it is concluded that the radial component of
geodesic deviation vector becomes zero while angular component
remain finite for certain initial condition.

\end{document}